\documentclass[10pt]{article}

\usepackage{epsfig}
\usepackage{amsfonts}
\usepackage{color}
\usepackage{amssymb}

\long\def\comment#1{}  

\long\def\comment#1{}
\newcommand{\ok}{\hfill $\Box$}

\newcommand{\ie}{\emph{i.e.\/}}
\newcommand{\eg}{\emph{e.g.\/}}

\newcommand{\bim}{\bar{I}}

\newcommand{\ans}{\mathit{ans}}
\newcommand{\epe}{\E_{\preceq}}

\newcommand{\indo}{\mathit{ind_S(o)}}

\newcommand{\obj}{\mathit{Obj}}
\newcommand{\pd}{\preceq_d}

\newcommand{\ACCA}{\mathcal{H}}

\newcommand{\C}{\mathcal{C}}
\newcommand{\E}{\mathcal{E}}
\newcommand{\EL}{\mathcal{L}}

\newcommand{\hato}{\mathcal{H}_o}

\newcommand{\N}{\mathcal{N}}

\newcommand{\PI}{\mathcal{P}}

\newcommand{\ES}{\mathcal{S}}

\newcommand{\V}{\mathcal{V}}
\newcommand{\s}{\sqsubseteq}
\newcommand{\ask}{\textsc{Ask}}
\newcommand{\askc}{\textsc{Ask}$_c$}
\newcommand{\askce}{\textsc{Ask}$_c^{ext}$}
\newcommand{\askca}{\textsc{Ask}$_c^{alt}$}
\newcommand{\query}{\textsc{Query}}
\newcommand{\qe}{\textsc{Qe}}
\newcommand{\tell}{\textsc{Tell}}
\newcommand{\tellc}{\textsc{Tell}$_c$}
\newcommand{\tellce}{\textsc{Tell}$_c^{ext}$}
\newcommand{\pid}{\emph{PID}}
\newcommand{\qid}{\emph{QID}}
\newcommand{\qp}{\emph{QP}}
\newcommand{\qidone}{\emph{QID}$_1$}
\newcommand{\qpone}{\emph{QP}$_1$}
\newcommand{\subp}{\mathit{SP}}

\newcommand{\bl}
  {\begin{list}{--}{
    \setlength{\topsep}{4pt}
    \setlength{\itemsep}{0pt}
    \setlength{\parsep}{4pt}
    \setlength{\partopsep}{0pt}}}
\newcommand{\el}{\end{list}}

\newtheorem{definition}{Definition}
\newtheorem{proposition}{Proposition}
\newtheorem{lemma}{Lemma}

\newtheorem{corollary}{Corollary}
\begin{document}

\title{Query Evaluation in P2P Systems of Taxonomy-based Sources: Algorithms,
Complexity, and Optimizations}

\author{Carlo Meghini \\
  Consiglio Nazionale delle Ricerche \\
  Istituto della Scienza e delle Tecnologie della Informazione \\
  Pisa, Italy\\
  {\tt carlo.meghini@isti.cnr.it} \and
  Yannis Tzitzikas \\ Department of Computer Science \\ University of
  Crete, Heraklion, Greece \\ Institute of Computer Science \\ Foundation for Research
  and Technology -- Hellas (FORTH-ICS)\\
  {\tt tzitzik@ics.forth.gr} \and
  Anastasia Analyti \\ Institute of Computer Science \\ Foundation for Research
  and Technology -- Hellas (FORTH-ICS)\\ Crete, Greece\\
  {\tt analyti@ics.forth.gr}} 

\date{}
\maketitle

\begin{abstract}
  In this study we address the problem of answering queries over a peer-to-peer
  system of taxonomy-based sources.  A taxonomy states subsumption relationships
  between negation-free DNF formulas on terms and negation-free conjunctions of
  terms.  To the end of laying the foundations of our study, we first consider
  the centralized case, deriving the complexity of the decision problem and of
  query evaluation.  We conclude by presenting an algorithm that is efficient in
  data complexity and is based on hypergraphs.  More expressive forms of
  taxonomies are also investigated which however lead to intractability.  We
  then move to the distributed case, and introduce a logical model of a network
  of taxonomy-based sources. On such network, a distributed version of the
  centralized algorithm is then presented, based on a message passing paradigm,
  and its correctness is proved.  We finally discuss optimization issues, and
  relate our work to the literature.
\end{abstract}

\section{Introduction}
\label{sec:intro}

Consider a tetrad $(T, \preceq, \obj, I)$ where $T$ is a set of terms, $\preceq$
is a subsumption relation over concepts expressed using $T$ (e.g.
$\mathtt{(Animal \wedge FlyingObject}) \vee \mathtt{Penguin} \preceq
\mathtt{Bird}$), $\obj$ is a set of objects and $I$ is a function from $T$ to
$\PI(\obj)$, assigning a description (\ie, a set of terms) to each object.  Now
assume that all these are not stored at a single place but they are distributed
over a set $\N=\{\ES_1, \ldots, \ES_n\}$ of independent peers.  Moreover assume
that each peer $S_i$ can have zero, one or more $\preceq$-relationships between
its terms (i.e.  $T_i$) and some concepts over the terminologies of other peers
(e.g.  $\mathtt{Parrot_j \preceq Birds_i}$ and $\mathtt{Animal_k \wedge Flying_k
  \preceq Birds_i} $).  In this paper we address the problem of answering
Boolean queries over this kind of systems.

Some parts of the work reported in this paper have been already published.
Namely,~\cite{TzitzikasMeghiniER03} presents a first model of a network of
articulated sources, while \cite{TzitzikasMeghiniCoopIS03} studies query
evaluation on taxonomies including only term-to-term subsumption relationships.
Finally,~\cite{ArticulatedODBASE04} presents a procedure for evaluating queries
over centralized sources supporting term-to-query subsumption relationships, as
well as hardness results for extensions.  In this paper,
\begin{itemize}
\item we consider from the start the most complex type of subsumption for which
  we can propose an efficient query evaluation procedure, allowing subsumption
  relationships between negation-free DNF combinations of terms and
  negation-free conjunctions of terms. We then place the hardness results
  presented in~\cite{ArticulatedODBASE04} in context, thus showing that any
  Boolean extension of the expressive power of subsumption leads to
  intractability of the query answering problem;
\item we ground the centralized query evaluation procedure for this kind of
  sources, presented in~\cite{ArticulatedODBASE04}, on solid theoretical basis,
  proving its correctness, and linking it to the existing algorithmic and
  complexity literature;
\item we present a distributed query evaluation procedure, based on a functional
  model of a peer; correctness and complexity of this procedure are given;
\item we describe optimization techniques that can be used for improving the
  efficiency of query evaluation;
\item we relate our work to the existing literature on peer-to-peer systems.
\end{itemize}

The paper is structured as follows: Section \ref{sec:bg} gives the background on
peer-to-peer systems, while Section \ref{sec:exts} introduces sources,
presenting the centralized query evaluation procedure.  Networks of sources are
considered in Section \ref{sec:Mediators}, where our algorithm for query
evaluation on networks is presented,
and
Section \ref{sec:opt} discusses optimization issues.
Section \ref{sec:rw} compares our work
with related work and Section \ref{sec:conc} concludes the paper.

\section{Background}
\label{sec:bg}

A peer-to-peer (P2P) system is a distributed system in which
participants (the peers) rely on one another for service, rather
than solely relying on dedicated and often centralized servers. The
most popular P2P systems have focused on specific application
domains like music file sharing \cite{Napster01,Gnutella,Kazaa}) or
on providing file-system-like capabilities \cite{Bolosky00}. In most
of the cases,  these systems do not provide semantic-based retrieval
services as the name of an object (e.g. the title of a music file)
is the only means for describing the contents of an object.

Semantic-based retrieval in P2P systems is a great challenge that
raises questions about data models, conceptual modeling, query
languages, algorithms and data structures for query evaluation, and
techniques for  dynamic schema mapping. Roughly, the language that
can be used for indexing the objects of the domain and for
formulating semantic-based queries, can be {\em free} (e.g natural
language) or {\em controlled}, i.e. object descriptions and queries
may have to conform to a specific vocabulary and  syntax. The former
case, resembles distributed Information Retrieval (IR)  systems and
this approach is applicable in the case where the objects of the
domain have a textual content (e.g.
\cite{LingWISE02,Koubarakis02,planetP2003,pSearch2002}). In the
latter case,  the objects of a peer are indexed according to a
specific conceptual model represented in a particular data model
(e.g. relational, object-oriented, logic-based, etc), and  content
searches are formulated using a specific query language. Of course,
a P2P system might impose a single conceptual model on all
participants to enforce uniform, global access, but this will be too
restrictive.  Alternatively, a limited number of conceptual models
may be allowed, so that traditional information mediation and
integration techniques will likely apply (with the restriction that
there is no central authority), e.g. see
\cite{Edutella02,Edutella02b}.

The case of fully heterogeneous conceptual models makes uniform
global access extremely challenging and this  is the focus of this
paper.
From a data modeling point of view several approaches for P2P
systems have been proposed recently, including relational-based
approaches \cite{Bernstein02}, XML-based approaches \cite{Halevy03b}
and RDF-based \cite{Edutella02b}.


In this paper we consider the  fully heterogeneous conceptual model
approach (where each peer can have its own schema), with the only
restriction that each conceptual model is represented as a taxonomy.
A taxonomy can
range from
a simple tree-structured hierarchy of terms,
to the concept lattice derived by Formal Concept Analysis
\cite{Ganter99}, or to the concept lattice of a Description Logics
theory. \comment{
        This  taxonomy-based conceptual modeling approach
        has three main advantages (for more see \cite{TzitzikasMeghiniER03}): (a) it is
        very easy to create the conceptual model of a source, (b) the integration of
        information from multiple sources can be done easily, and
        (c) automatic articulation using data-driven methods (like the one presented in
        \cite{TzitzikasMeghiniCIA03}) are possible\footnote{
            We do not need to make the assumption that
            there is a set Objects (each element of which has the same identify in the entire network).
            It is good for conjunctions over different peers and see below.
        }.
==} Specifically, according to our model, each peer consists of a
{\em taxonomy}, an {\em object base}, i.e. a database that contains
descriptions of the objects according to the taxonomy, and
a number of (one-way) {\em articulations} to some of the other peers
of the network, where an articulation is actually a  mapping between
terms of the peer  and terms (or queries) of other peers.
Articulations aim at bridging the inevitable naming, granularity and
contextual heterogeneities that     may exist between the taxonomies
of the peers (for some examples see  \cite{TzitzikasMeghiniER03}).
For example,
the taxonomy of a peer $\ES_1$ could be the following:
    $\{~
    \mathtt{Penguin} \preceq \mathtt{Animal}$,
    $\mathtt{Pelican} \preceq \mathtt{Animal}$,
    $\mathtt{Ostrich} \preceq \mathtt{Animal}$,
    $\mathtt{(Animal \wedge FlyingObject}) \vee \mathtt{Penguin} \vee \mathtt{Ostrich} \preceq \mathtt{Bird} ~\}$.
The object base of $\ES_1$ could be the following:
    $\{~
     \mathtt{Ostrich}(1)$,
    $\mathtt{Bird}(2)$,
    $\mathtt{Animal}(3)$,
    $\mathtt{FlyingObject}(3) ~\}$.
$\ES_1$ could have an  articulation to a peer $\ES_2$ like
$\{~
  \Pi\iota\nu\gamma\kappa o\upsilon \acute{\iota}\nu o\varsigma_2 \preceq \mathtt{Penguin}$,
$ \Pi\epsilon\lambda\epsilon\kappa\acute{\alpha} \nu o\varsigma_2 \preceq \mathtt{Pelican}  ~\}$,
an articulation to a peer $\ES_3$ like
$\{~ \mathtt{Animale}_3 \wedge \mathtt{Alato}_3 \preceq \mathtt{Birds} ~\}$,
and  an articulation to two peers $\ES_4, \ES_5$ of the form:
$\{~ (\mathtt{Fliegentier}_4) \vee (\mathtt{Animal}_5 \wedge \mathtt{Volant}_5) \preceq
(\mathtt{Animal} \wedge \mathtt{FlyingObject}) ~\}$.

The articulations can be exploited for finding objects  in the
network using content-based queries, for publishing objects and
their descriptions to the network, and for obtaining more rich
descriptions of the objects (by aggregating their descriptions
according to different conceptual models)\footnote{
    The latter is possible only if the objects have a unique global identity
    in the entire network (like URI for example).
}.
Apart from determining query propagation, these mappings are
actually used for translating the query into a vocabulary that the
recipient can understand (and thus answer). In certain cases, these
inter-taxonomy mappings could be constructed automatically (e.g.
using the  data-driven method proposed in
\cite{TzitzikasMeghiniCIA03}).

 \comment{==== CIA 2003 paper
    This method, that is called {\em ostensive},
    is possible if the objects of the domain  have a unique global identity.
    It is called {\em ostensive} because the meaning of each term is explained by ostension,
    i.e. by pointing to something (here, to a set of objects) to which the term applies.
    For example, the word "rose" can be defined ostensively by pointing to a rose and saying "that is a rose".
    Instead, the verbal methods of term definition (e.g. the  synonyms or the  analytic method)
    presuppose that the learner already knows some other terms and, thus, they are useless
    to someone who does not know these terms; e.g. verbal word definitions  are useless to a
    small child   who has not learnt any words at all.
    If each peer that joins the network is obliged to index   a given reference collection of objects,
    when this would allow running this algorithm
    (even in cases where the objects stored at the peers are mutually disjoint).
====}


The placement of our work with respect to other logic-founded approaches for
query evaluation over P2P systems
is given in Section \ref{sec:rw}.


\section{Information sources}
\label{sec:exts}

This Section defines information sources and derives algorithms and complexity
results for querying them. These results will be applied later, upon studying
networks of sources. The model is first introduced; the computational and
algorithmic foundations of the query evaluation problem are then given;
Section~\ref{sec:psqees} presents an efficient query evaluation method. Finally,
three extensions of information sources are discussed: those having negation in the
taxonomy, those having negation only in the query language, and those having
disjunction in the taxonomy. For all these, the query evaluation problem is
studied, deriving complexity results or correct and efficient algorithms, if
any.

\subsection{The model}
\label{sec:em}

The basic notion of the model is that of \emph{terminology:} a terminology $T$
is a non-empty set of terms. A terminology comes with an associated language for
constructing more complex terms, called \emph{queries,} from the given ones.

\begin{definition}[Query] \label{def:QueryLang}
  \emph{The \emph{query language} associated to a terminology $T,$ $\EL_T,$ is
    the language defined by the following grammar, where $t$ is a term of $T:$
    \begin{tabbing}
      ind \= m \= ::= \= \kill
      \> $q$ \> ::= \> $d ~ | ~ q \vee d$ \\
      \> $d$ \> ::= \> $t ~ | ~ t \wedge d.$
    \end{tabbing}
    An instance of $q$ is called a \emph{query,} while an instance of $d$ is
    called a \emph{conjunctive query}. Each $d$ component of a query $q$ is called {\em disjunct} of $q$}. \ok
\end{definition}

Terms and queries can be used for defining taxonomies.

\begin{definition}[Taxonomy] \label{def:Tax}
  \emph{A \emph{taxonomy} is a pair $(T, \preceq)$ where $T$ is a terminology
    and $\preceq$ is a binary relation between queries, $\preceq \; \subseteq (\EL_T\times
    \EL_T),$ which is reflexive and transitive, such that $q\preceq q'$ and
    $q\not=q'$ imply that $q'$ is a conjunctive query.} \ok
\end{definition}

If $(q,q') \in \; \preceq,$ we say that $q$ is subsumed by $q'$ and we write $q\preceq
q'.$ The reason for having only conjunctive queries as right-hand sides of
non-trivial subsumption relationships is computational, and will be discussed
later.

\begin{definition}[Interpretation] \label{def:Int}
  \emph{An \emph{interpretation} for a terminology $T$ is a pair $(\obj,I)$,
    where $\obj$ is a finite set of objects and $I$ is a total function $I: T
    \rightarrow \PI(\obj).$ } \ok
\end{definition}

Interpretations can be extended to queries in an intuitive way, thus defining
the semantics of the query language:

\begin{definition}[Query extension] \label{def:QExt}
  \emph{Given an interpretation $I$ of a terminology $T$ and a query $q\in
    \EL_T,$ the \emph{extension of q in I,} $q^I,$ is defined as follows:
  \begin{enumerate}
  \item $(q\vee d)^I=~q^I\cup~ d^I$
  \item $(d\wedge t)^I=~d^I\cap~ t^I$
  \item $t^I=I(t).$ \ok
  \end{enumerate}}
\end{definition}

Since the function $\cdot^I$ is an extension of the interpretation function $I,$
we will simplify notation and will write $I(q)$ in place of the formally correct
$q^I.$ We can now define a taxonomy-based source, called \emph{information
  source} or simply \emph{source.}

\begin{definition}[Information source] \label{def:IS}
  \emph{An \emph{information source} $S$ is a 4-tuple
    $S=(T_S,\preceq_S,\obj_S,I_S)$, where $(T_S,\preceq_S)$ is a taxonomy and 
    $(\obj_S,I_S)$ is an interpretation for $T_S.$ } \ok
\end{definition}

When no ambiguity will arise, we will simplify notation by omitting the
subscript in the components of sources. In addition, an interpretation will be
equated with its interpretation function $I.$ Given a source $S=(T,
\preceq,\obj,I)$ and an object $o\in\obj,$ the \emph{index of o in S,} $\indo,$
is given by the terms in whose interpretation $o$ belongs, \ie:
\[
\indo=\{t\in T~|~o\in I(t)\}.
\]
\indent
Some interpretations better reflect the semantics of subsumption.

\begin{definition}[Models of a source] \label{def:Models}
  \emph{Given two interpretations $I$, $I'$ of the same terminology $T,$
    \begin{itemize}
    \item $I$ is a {\em model} of the taxonomy $(T,\preceq)$ if $q \preceq q'$
      implies $I(q) \subseteq I(q');$
    \item $I$ is smaller than $I',$ $I \leq I'$, if $I(t) \subseteq I'(t)$ for
      each term $t \in T;$
    \item $I$ is a {\em model} of a source $S=(T,\preceq,\obj,I')$ if it is a
      model of $(T,\preceq)$ and $I'\leq I.$ \ok
    \end{itemize}
    }
\end{definition}

The notion of model of a source can be used to obtain a simpler, but equivalent,
notion of source, in which (non-trivial) subsumption relationships relate
conjunctive queries to terms. The equivalence is based on the observation that
the propositional formula:
\[
(C_1\vee\ldots\vee C_n) \rightarrow (t_1\wedge\ldots\wedge t_m)
\]
where each $C_i$ in the left hand-side is any propositional formula, is
logically equivalent to the formula:
\[
(C_1 \rightarrow t_1) \wedge (C_1 \rightarrow t_2) \wedge \ldots \wedge (C_1
\rightarrow t_m) \wedge \ldots \wedge (C_n \rightarrow t_1) \wedge (C_n
\rightarrow t_2) \wedge \ldots \wedge (C_n \rightarrow t_m),
\]
that is, the two formulae have the same models. Formally, the
\emph{simplification} of a taxonomy $(T,\preceq),$ is the taxonomy
$(T,\sigma(\preceq)),$ where $\sigma(\preceq)$ is the reflexive and transitive
closure of the following relation\footnote{The transitive reduction of a binary
  relation $R$ on a set $X,$ is defined as~\cite{fejer} $R^r=R_1 \setminus
  R_1^2,$ where $R_1=R\setminus\{(a,a)~|~a\in X\}$ and $R_1^2=R_1\circ R_1.$ In
  practice, $R^r$ is $R$ without reflexive and transitive relationships, and its
  graphical rendering is generally known as the \emph{Hasse diagram} of $R.$}:
\[
\{(C,t)~|~(C_1\vee\ldots\vee C_n,t_1\wedge\ldots\wedge
t_m)\in\preceq^r,~C\in\{C_1,\ldots,C_n\}, ~t\in\{t_1,\ldots,t_m\}\}.
\]
Correspondingly, the simplification of a source $S=(T,\preceq,\obj,I)$ is the
source $\sigma(S)=(T,\sigma(\preceq),\obj,I).$ It is not difficult to see that:

\begin{proposition} \label{prop:red}
  \emph{$J$ is a model of a source $S$ if and only if it is a model of
    $\sigma(S).$ } \ok
\end{proposition}

Based on the last Proposition, from now on we will use the terms ``taxonomy''
and ``source'' as synonyms of ``simplified taxonomy'' and ``simplified source'',
respectively. Formally, $(T,\preceq)$ and $S$ will stand for
$(T,\sigma(\preceq))$ and $\sigma(S),$ respectively.

A second usage of the notion of model is to define the query-answering function
$\ans$ on sources.

\begin{definition}[Answer] \label{def:AnswerQS}
  \emph{Given a source $S=(T,\preceq,\obj,I)$ and a query $q\in\EL_T,$ the
    \emph{answer of q in S,} $\ans(q,S),$ is given by $\ans(q,S)=\{o\in
    \obj~|~o\in J(q) \mbox{ for all models $J$ of $S$}\}.$ } \ok
\end{definition}

Indeed, we only need to consider term queries, because non-term queries can be
embedded in the taxonomy. Specifically:

\begin{proposition} \label{prop:termqonly}
  \emph{For all sources $S=(T,\preceq,\obj,I)$ and non-term queries
    $q\in\EL_T,$ let $t_q\not\in T$ and
    \begin{eqnarray*}
      T^q & = & T\cup\{t_q\} \\
      (\preceq^q)^r & = & \preceq^r \cup\{(t_1\wedge\ldots\wedge t_m,t_q)
      |~t_1\wedge\ldots\wedge t_m \mbox{ is a disjunct of }q\} \\
      I^q & = & I \cup \{(t_q,\emptyset)\}.
    \end{eqnarray*}
    Then, $\ans(q,S)=\ans(t_q,S^q)$ where $S^q=(T^q, \preceq^q,\obj,I^q).$ }
  \ok
\end{proposition}
In practice, the terminology $T^q$ includes one additional term $t_q,$ which has
an empty interpretation and subsumes each query disjunct $t_1\wedge\ldots\wedge
t_m$ of $q$. The size of $S^q$ is clearly polynomial in the size of $S$
and $q.$

In light of the last Proposition, the problem of query evaluation amounts to
determine $\ans(t,S)$ for given term $t$ and source $S,$ while the corresponding
decision problem consists in checking whether $o\in\ans(t,S),$ for a given
object $o.$

Query evaluation is strictly related to the unique minimal model of a source.

\begin{proposition} \label{prop:MinModel}
  \emph{For all sources $S=(T,\preceq,\obj,I)$ and terms $t\in T,$
    the unique minimal model of $S,$ $\bim,$ is given by
  \[
  \bim(t) = \bigcup \{I(u)~|~u\in T,~u \preceq t\}~\cup~\bigcup \{ \bim(q) \; |
  \; q \preceq t, \; q = t_1\wedge\ldots\wedge t_m,~m > 1 \}.
  \]
  Moreover, $ans(t,S)=\bim(t).$
} \ok
\end{proposition}

\subsection{Foundations}
\label{sec:fqees}

In this Section, we consider the computational foundations of query evaluation,
starting from those of the more fundamental decision problem.

\subsubsection{The decision problem}
\label{sec:dec}

Given a source $S=(T,\preceq,\obj,I)$, $o \in \obj$, and $t \in T$, the decision problem
$o \in \ans(t,S)$ is P-complete in the size of the taxonomy. The hardness
part of the proof is based on the following polynomial time reduction from the
decision problem $P\models A$ in propositional datalog, known to be
P-complete~\cite{degv01}:
\begin{itemize}
\item the terminology $T$ is given by the letters occurring in $P;$
\item $\preceq$ is the reflexive and transitive closure of the binary relation,
  defined as follows:
  \[
  \{(t_1\wedge\ldots\wedge t_m,t_0)~|~t_0 \leftarrow t_1,\ldots,t_m\in P\}.
  \]
\item $\obj=\{1\};$
\item the interpretation function $I$ is defined as follows: for each term $t_0 \in
  T,$
  \[
  I(t_0)=\left\{
    \begin{array}{ll}
      \{1\}     & \mbox{if }t_0\leftarrow~ \in P \\
      \emptyset & \mbox{otherwise}
    \end{array}
    \right.
    \]
\end{itemize}
It is easy to see that $P\models A$ if and only if
$1\in\ans(A,(T,\preceq,\obj,I)),$ thus obtaining hardness in the size of the
program $P.$

For the membership part of the proof, we rely on an opposite reduction, which
will also be used later. Let $S=(T,\preceq,\obj,I)$ be a source, $o\in\obj$ and
$t\in T.$ Define $P_S$ to be the following propositional datalog program:
\[
P_S = C_S \cup I_S \cup Q_S
\]
where
\begin{eqnarray*}
  C_S & = & \{t_0\leftarrow t_1,\ldots,t_m~|~(t_1\wedge\ldots\wedge
  t_m,t_0)\in\preceq^r \} \\
  I_S & = & \{u\leftarrow ~|~u\in\indo\} \\
  Q_S & = & \{\leftarrow t\}
\end{eqnarray*}
The size of $P_S$ is polynomial in the size of the taxonomy.
It is easy to see that:

\begin{lemma}\label{lemma:exts}
  \emph{For all sources $S=(T,\preceq,\obj,I),$ $o\in\obj$ and $t\in T,$
    $o\in\ans(t,S)$ iff $P_S$ is unsatisfiable. } \ok
\end{lemma}
This proves the membership of the decision problem in P, hence its
P-completeness.  From the P-completeness in the size of the taxonomy of the
decision problem, the P-completeness in the size of the information source\footnote{
The {\em size} of an information source comprises the size of its taxonomy and the size of its interpretation, 
i.e., what is called combined complexity, in the database literature.}
 of
the query evaluation problem follows.

From an algorithmic point of view, the decision problem relies on directed
B-hypergraphs, which are introduced next. We will mainly use definitions and
results from~\cite{gallo93}.

A \emph{directed hypergraph} is a pair $\ACCA=(\V,\E),$ where $\V=\{v_1,
v_2,\ldots, v_n\}$ is the set of vertices and $\E=\{E_1, E_2,\ldots, E_m\}$ is
the set of directed hyperedges, where $E_i=(\tau(E_i),\chi(E_i))$ with
$\tau(E_i),\chi(E_i)\subseteq \V$ for $1\leq i\leq m.$ $\tau(E_i)$ is said to be
the \emph{tail} of $E_i,$ while $\chi(E_i)$ is said to be the \emph{head} of
$E_i.$ A \emph{directed B-hypergraph} (or simply \emph{B-graph}) is a directed
hypergraph, where the head of each hyperedge $E_i,$ denoted as $h(E_i),$ is a
single vertex.

\begin{figure}[htbp]
  \centering
  \begin{minipage}{7cm}
    \begin{picture}(5.5,2.5)
      \put(1,2){\makebox(0.5,0.5){$a1$}}
      \put(0.5,1){\makebox(0.5,0.5){$a2$}}
      \put(1.5,1){\makebox(0.5,0.5){$a3$}}
      \put(1,0){\makebox(1,0.5){$b1\wedge b2$}}
      \put(3.25,0){\makebox(1,0.5){$b1\wedge b3$}}
      \put(0,0){\makebox(0.5,0.5){$b3$}}
      \put(3,2){\makebox(0.5,0.5){$b1$}}
      \put(4.75,2){\makebox(0.5,0.5){$b2$}}
      \put(2.5,1){\makebox(0.5,0.5){$c1$}}
      \put(3.5,1){\makebox(0.5,0.5){$c2$}}
      \put(4.5,1){\makebox(1,0.5){$c2\wedge c3$}}

      \multiput(0.75,1.5)(2,0){2}{\vector(1,2){0.25}}
      \multiput(1.75,1.5)(2,0){2}{\vector(-1,2){0.25}}
      \multiput(3.75,0.5)(1.25,1){2}{\vector(0,1){0.5}}
      \put(0.25,0.5){\vector(1,2){0.25}}
      \put(1.5,0.5){\vector(-1,1){0.5}}
    \end{picture}
  \end{minipage}
  \begin{minipage}{5.75cm}
    \begin{picture}(5.75,3.5)
      \put(0,1.5){\makebox(0.5,0.5){$a1$}}
      \put(1.5,1.5){\makebox(0.5,0.5){$a2$}}
      \put(3,1.5){\makebox(0.5,0.5){$b2$}}
      \put(4.5,1.5){\makebox(0.5,0.5){$c2$}}
      \put(1.5,2.5){\makebox(0.5,0.5){$a3$}}
      \put(3,2.5){\makebox(0.5,0.5){$b3$}}
      \put(4.5,2.5){\makebox(0.5,0.5){$c3$}}
      \put(3,0.5){\makebox(0.5,0.5){$b1$}}
      \put(4.5,0.5){\makebox(0.5,0.5){$c1$}}
      \multiput(1.5,2.5)(1.5,0){2}{\vector(-2,-1){1}}
      \multiput(1.5,1.75)(1.5,0){3}{\vector(-1,0){1}}
      \put(3,1){\line(-1,2){0.37}}
      \put(4.5,1.5){\vector(-2,-1){1}}
      \put(4.5,2.5){\line(-1,-2){0.38}}
      \put(4.5,0.75){\vector(-1,0){1}}
      \put(4.25,3){\oval(2,.75)[t]}
      \put(5.25,3){\line(0,-1){1.25}}
      \put(5.25,1.75){\vector(-1,0){.25}}
      \put(4.5,0.5){\oval(2.5,.75)[b]}
      \put(5.25,1.25){\oval(1,1)[tr]}
      \put(5.75,0.5){\line(0,1){.75}}
    \end{picture}
  \end{minipage}
\caption{A taxonomy and its B-graph}
  \label{fig:extsource}
\end{figure}
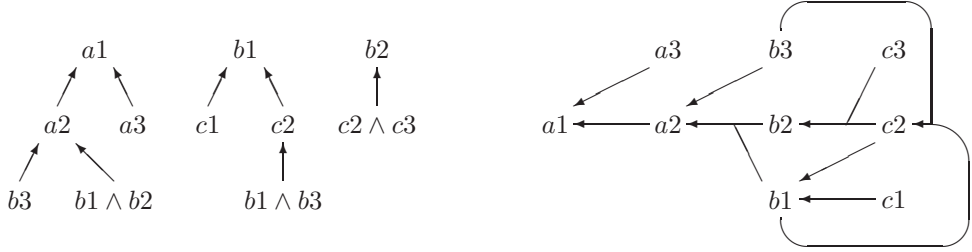

A taxonomy can naturally be represented as a B-graph whose hyperedges represent
one-to-one the subsumption relationships of the transitive reduction of the taxonomy.
In particular, the \emph{taxonomy
  B-graph} of a taxonomy $(T,\preceq)$ is the B-graph $\ACCA=(T,\epe)$
where
\[
\epe = \{(\{t_1,\ldots,t_m\},u)~|~(t_1\wedge\ldots\wedge t_m,u)\in\preceq^r\}
\]
Figure~\ref{fig:extsource} left presents a taxonomy, whose B-graph is shown in
the same Figure right.

A \emph{path} $P_{st}$ of length $q$ in a B-graph $\ACCA=(\V,\E)$ is a sequence
of nodes and hyperedges
\[
P_{st}=(s=v_1,E_{i_1},v_2,E_{i_2},\ldots,E_{i_q},v_{q+1}=t)
\]
where: $s\in\tau(E_{i_1}),$ $h(E_{i_q})=t$ and
$h(E_{i_{j-1}})=v_j\in\tau(E_{i_j})$ for $2\leq j\leq q.$ If $P_{st}$ exists,
$t$ is said to be \emph{connected} to $s.$ If $t\in\tau(E_{i_1}),$ $P_{st}$ is
said to be a \emph{cycle}; if all hyperedges in $P_{st}$ are distinct, $P_{st}$
is said to be \emph{simple.} A simple path is \emph{elementary} if all its
vertices are distinct.

A \emph{B-path} $\pi_{st}$ in a B-graph $\ACCA=(\V,\E)$ is a minimal (with respect to
deletion of vertices and hyperedges) hypergraph $\ACCA_\pi=(\V_\pi, \E_\pi),$
such that:
\begin{enumerate}
\item $\E_\pi\subseteq\E$
\item $\{s,t\}\subseteq\V_\pi$
\item $x\in\V_\pi$ and $x\neq s$ imply that $x$ is connected to $s$ in
  $\ACCA_\pi$ by means of a cycle-free simple path.
\end{enumerate}
Vertex $y$ is said to be \emph{B-connected} to vertex $x$ if a B-path $\pi_{xy}$
exists in $\ACCA.$

B-graphs and satisfiability of propositional Horn clauses are strictly
related. The B-graph {\em associated to} a set of Horn clauses has 3 types of
directed hyperedges to represent each clause:

\begin{itemize}
\item the clause $p \leftarrow q_1\wedge q_2\wedge \ldots\wedge q_s$ is represented by the hyperedge
  $(\{q_1,q_2, \ldots, q_s\}, p);$
\item the clause $ \leftarrow q_1\wedge q_2\wedge \ldots\wedge q_s$ is represented by the hyperedge
  $(\{q_1,q_2, \ldots, q_s\}, \mathit{false});$
\item the clause $p \leftarrow $ is represented by the hyperedge
  $(\{\mathit{true}\}, p).$
\end{itemize}
The following result is well-known:
\begin{proposition}[\cite{gallo93}] \label{prop:sat}
  \emph{A set of propositional Horn clauses is satisfiable if and only if in the
    associated B-graph, \emph{false} is not B-connected to \emph{true}.  } \ok
\end{proposition}

\begin{figure}[t]
  \centering
  \begin{picture}(6.5,3.5)

    \put(0,1.5){\makebox(0.5,0.5){$a1$}}
    \put(1.5,1.5){\makebox(0.5,0.5){$a2$}}
    \put(3,1.5){\makebox(0.5,0.5){$b2$}}
    \put(4.5,1.5){\makebox(0.5,0.5){$c2$}}
    \put(1.5,2.5){\makebox(0.5,0.5){$a3$}}
    \put(3,2.5){\makebox(0.5,0.5){$b3$}}
    \put(4.5,2.5){\makebox(0.5,0.5){$c3$}}
    \put(3,0.5){\makebox(0.5,0.5){$b1$}}
    \put(4.5,0.5){\makebox(0.5,0.5){$c1$}}
    \put(6,2.5){\makebox(0.5,0.5){\emph{true}}}


    \multiput(1.5,2.5)(1.5,0){2}{\vector(-2,-1){1}}
    \multiput(1.5,1.75)(1.5,0){3}{\vector(-1,0){1}}
    \put(3,1){\line(-1,2){0.37}}
    \put(4.5,1.5){\vector(-2,-1){1}}
    \put(4.5,2.5){\line(-1,-2){0.38}}
    \put(4.5,0.75){\vector(-1,0){1}}
    \put(5.8,2.75){\vector(-1,0){0.8}}
    \put(6,2.5){\vector(-2,-1){1}}
    \put(6,2.5){\vector(-2,-3){1}}


    \put(4.25,3){\oval(2,.75)[t]}
    \put(5.25,3){\line(0,-1){1.25}}
    \put(5.25,1.75){\vector(-1,0){.25}}
    \put(4.5,0.5){\oval(2.5,.75)[b]}
    \put(5.25,1.25){\oval(1,1)[tr]}
    \put(5.75,0.5){\line(0,1){.75}}

  \end{picture}
  \caption{An object graph}
  \label{fig:accaaqo}
\end{figure}
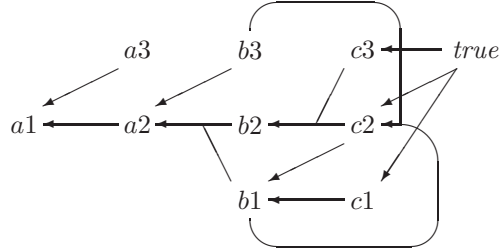

We now proceed to show the role played by B-connection in query evaluation. For
a source $S=(T,\preceq,\obj,I)$ and an object $o\in\obj,$ the \emph{object
  decision graph} (simply the \emph{object graph}) is the B-graph
$\ACCA_o=(T,\E_o),$ where
\[
\E_o = \epe \cup \bigcup\{(\{\mathit{true}\},u)~|~u\in\indo\}.
\]
Figure~\ref{fig:accaaqo} presents the object graph for the taxonomy shown in
Figure~\ref{fig:extsource} and an object $o$ such that $\indo=\{c1,c2,c3\}.$

We can now prove:

\begin{proposition} \label{prop:poqees}
  \emph{For all sources $S=(T,\preceq,\obj,I),$ terms $t\in T,$ and objects
    $o\in\obj,$ $o\in\ans(t,S)$ iff \emph{t} is B-connected to
    \emph{true} in the object graph $\hato.$} \\ 
    Proof: From Lemma~\ref{lemma:exts}, $o\in\ans(t,S)$ iff $P_S$ is
    unsatisfiable iff (by Proposition~\ref{prop:sat}) \emph{false} is
    B-connected to \emph{true} in the associated B-graph. By construction,
    $\hato$ is the B-graph associated to $P_S,$ where \emph{t} plays the role
    of \emph{false.} \ok
\end{proposition}

\subsubsection{Foundation of query evaluation}
\label{sec:answerSet}
The basic reason why the decision problem can be efficiently solved, is that it
requires traversing any hyperedge of the taxonomy B-graph at most once. In other
words, when deciding membership of an object to a query answer, any
(non-trivial) subsumption relationship needs to be used no more than once.
However, this is not the case for query evaluation, for in this case all objects
must be considered at once as potential candidates for the answer, and therefore
a hyperedge can be traversed more than once, in different ways. From a more
technical point of view, in deciding whether $o\in\ans(t,S),$ we consider the
cycle-free simple paths from any term in $\indo$ to $t.$ These paths make up the
B-path $\ACCA_o.$ Instead, in computing $\ans(t,S),$ we need to consider a much
larger hypergraph, call it $\ACCA_\obj,$ in which \emph{true} is connected to
\emph{all} terms in $T$ that belong to at least one object index. $\ACCA_\obj$
is made up of all cycle-free simple paths from \emph{any} term to $t.$ Now, it
is not difficult to see that these paths may be exponentially many in the size
of the taxonomy. As an illustration, let us consider the taxonomy whose B-graph
contains the following hyperedges:

\begin{center}
\begin{tabular}{lllll}
  $h_1:(\{u_1,v_1\},u_2)$ & $h_2:(\{u_2,v_2\},u_3)$ & $h_3:(\{u_3,v_3\},u_4)$ &
  $h_4:(\{u_4,v_4\},u_5)$ & $h_5:(\{u_5,v_5\},t)$ \\
  $g_1:(\{u_1,v_1\},v_2)$ & $g_2:(\{u_2,v_2\},v_3)$ & $g_3:(\{u_3,v_3\},v_4)$ &
  $g_4:(\{u_4,v_4\},v_5)$ &
\end{tabular}
\end{center}
Let us assume $t$ is the query term. It is easy to verify that there are
$2^{4}$ cycle-free simple paths connecting $u_1$ to $t,$ one for each sequence of
the form
\begin{center}
  $(u_1~f_1~x_2~f_2~x_3~f_3~x_4~f_4~x_5~h_5~t)$
\end{center}
where $f_i$ can be either $h_i$ (in which case $x_{i+1}$ is $u_{i+1}$) or $g_i$
(in which case $x_{i+1}$ is $v_{i+1}$) for $1\leq i\leq 4.$ In fact, any object
$o$ whose index $\indo$ contains either both $u_j$ and $v_j$ (for some $1\leq
j\leq 5$) or $t,$ is in the answer of the query, and so there is an exponential
number of indices which qualify for the query. In order to avoid examining all
these indices, a smart query evaluation algorithm could try to generate only the
minimal ones, which in our case are just 6. However, finding all minimal
qualifying indices is an NP hard problem.

\begin{figure}[htbp]
  \centering
  \begin{picture}(10,3.5)(-4,0)

    \put(-4,1.5){\makebox(3,2)
      {
        \begin{minipage}{6cm}
          \begin{tabbing}
            $\C=\{$\=$\{a,b\},$ \\
            \> $\{b,c,d\},$ \\
            \> $\{b,c,e,f\}\}$
          \end{tabbing}
        \end{minipage}
      }}

    \put(0.5,1.5){\makebox(0.5,0.5){$u_1$}}
    \put(2,1.5){\makebox(0.5,0.5){$u_2$}}
    \put(3.5,1.5){\makebox(0.5,0.5){$u_3$}}

    \put(0,0){\makebox(0.5,0.5){\emph{a}}}
    \put(1,0){\makebox(0.5,0.5){\emph{b}}}
    \put(2,0){\makebox(0.5,0.5){\emph{c}}}
    \put(3,0){\makebox(0.5,0.5){\emph{d}}}
    \put(4,0){\makebox(0.5,0.5){\emph{e}}}
    \put(5,0){\makebox(0.5,0.5){\emph{f}}}
    \put(2,3){\makebox(0.5,0.5){\emph{t}}}

    \put(0.25,0.5){\vector(1,2){0.5}}
    \put(1.25,0.5){\vector(-1,2){.5}}
    \put(1.25,0.5){\vector(1,1){1}}
    \put(1.5,0.5){\vector(2,1){2}}
    \put(2.25,0.5){\vector(0,1){1}}
    \put(2.25,0.5){\vector(3,2){1.5}}
    \put(3.25,0.5){\vector(-1,1){1}}
    \put(4.25,0.5){\vector(-1,2){.5}}
    \put(5,0.5){\vector(-1,1){1}}

    \put(0.75,2){\line(3,1){1.5}}
    \put(2.25,2){\vector(0,1){1}}
    \put(3.75,2){\line(-3,1){1.5}}

  \end{picture}
  \caption{A collection of sets $\C$ and the B-graph of the corresponding
    taxonomy $(T_\C, \preceq_\C)$}
  \label{fig:GAC}
\end{figure}
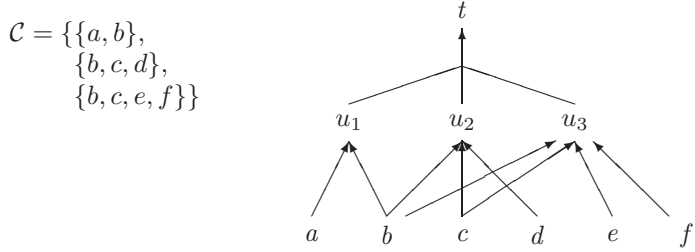

In proof, let us define an \emph{answer set A} for a term query $t$ to a source
$S=(T,\preceq,\obj,I),$ to be a set of terms $A\subseteq T,$ such that if the
index of an object $o,$ $\indo,$ has all the terms in $A,$ then $o$ is an answer
for $t$ in $S;$ formally, $A\subseteq\indo$ implies $o\in\ans(t,S).$ We now
present a polynomial time reduction from MINIMAL HITTING SET, a problem known to
be NP-complete, to the problem of finding a minimal answer set for $t$ in $S.$
We recall the notion of hitting set: Given a collection $\C$ of subsets of a set
$C,$ a \emph{hitting set} for $\C$ is a set $C'\subseteq C$ such that $C'$
contains at least one element from each subset in $\C.$ The basic working of the
reduction is exemplified in Figure~\ref{fig:GAC}, the left part of which
shows the collection $\C,$ while the right part shows the corresponding
taxonomy. The query is $t.$ In general, letting $\C=\{C_1,\ldots,C_k\}$ be a
collection of subsets of a set $C,$ the corresponding source
$S_\C=(T_\C,\preceq_\C,\emptyset,\emptyset)$ and term query $t_\C$ are defined
as follows:
\begin{itemize}
\item $T_\C=C\cup \{t,u_1,\ldots,u_k\}$ where $t\not\in C$ and $u_i\not\in
  C$ for all $1\leq i\leq k.$
\item $\preceq_\C^r=\bigcup_{1\leq j\leq k}
  \{(x,u_j)~|~x\in C_j\} \cup\{(u_1\wedge\ldots\wedge u_k,t)\}$
\item $t_\C=t.$
\end{itemize}
It can be easily proved that this is a polynomial time reduction and, of course, it holds that
$\preceq_\C$ is reflexive and transitive. Moreover, the terms from which each
term $u_i$ can be reached in the taxonomy B-graph are those of the i-th
collection in $\C,$ plus the element $u_i.$ Consequently, each hitting set for
$\C$ contains a sub-term of each $u_i,$ therefore it is an answer set for $t_\C$
and $S_\C;$ in addition, the minimality of the former implies that of the
latter. The converse is not true, because a minimal answer set $X$ for $t_\C$
and $S_\C$ may contain a ``foreign'' term $u_i.$ However, this is harmless, for
$u_i$ can be replaced in $X$ by any of its sub-terms and the result is still a
minimal hitting set for $\C.$ This proves the NP-hardness.

Notice that the reduction uses a much simpler type of information source than
the one we consider in the present study, namely one whose taxonomy has only one
hyperedge. Also, we have left the domain and the interpretation of $S_\C$ empty
in order to stress that they play no role in the reduction.

It is not difficult to prove membership of the problem in NP, from which the
NP-completeness in the size of the taxonomy of finding one minimal answer set
follows. However, query evaluation requires finding \emph{all} minimal answer
sets, thus the complexity of this latter problem is much worse, in fact we
believe that it is PSPACE-complete.

We now turn to the derivation of an algorithm for query evaluation, whose
complexity is polynomial in the size of the information source (which may be
exponentially higher than that of the taxonomy, of course).

\subsection{Query evaluation}
\label{sec:psqees}

Proposition~\ref{prop:MinModel} does not directly lead to a simple method for
query evaluation, as it may yield a recursive set of equations. As an
illustration, let us consider the query $b1$ in our example source. We have:
\begin{eqnarray*}
  \bim(b1) & = & I(b1)\cup I(c1)\cup I(c2)\cup\bim(b1\wedge b3) \\
  \bim(b1\wedge b3) & = & \bim(b1)\cap\bim(b2).
\end{eqnarray*}
The standard datalog approach to solve this problem is to map the program into a
system of equations on relations, which is then solved by applying an iterative
method (see Chapter 13 of \cite{AbHV95}). Given the simplified form of datalog programs that we are dealing with,
we propose a simpler method to perform query evaluation, based on B-graphs. Our
method relies on the following result, which is just a re-phrasing of
Proposition~\ref{prop:poqees}:

\begin{corollary} \label{prop:iteli}
  \emph{For all sources $S=(T,\preceq,\obj,I),$ $o\in\obj$ and term queries
    $t\in T,$ $o\in\ans(t,S)$ if and only if either $o\in I(t)$ or there exists
    a hyperedge $(\{u_1,\ldots,u_r\},t)\in\epe$ such that $o\in\bigcap
    \{\ans(u_i,S)~|~1\leq i\leq r\}.$} \ok
\end{corollary}
This corollary simply ``breaks down'' Proposition~\ref{prop:poqees} based on the
distance between $t$ and \emph{true} in the object graph $\hato.$ If $o\in
I(t),$ then $t\in\indo,$ hence there is a hyperedge (in fact, a simple arc) from
\emph{true} to $t$ in $\hato,$ which are 1 hyperedge distant from each other. If
$o\not\in I(t),$ then there are at least two hyperedges in between \emph{true}
and $t.$ Let us assume that $h$ is the one whose head is $t.$ Since
$t$ is B-connected to \emph{true,} each term $u_i$ in the tail of $h$ is
B-connected to \emph{true.} But this simply means, again by
Proposition~\ref{prop:poqees}, that $o\in\ans(u_i,S)$ for all the terms $u_i,$
and so we have the Corollary. Notice that, by point 3 in the definition of
B-path, $t$ is connected to each $u_i$ by a cycle-free simple path; this fact is
used by the procedure \qe~in order to correctly terminate in presence of loops in
the taxonomy B-graph $\ACCA.$

\begin{figure}
{\small
\begin{tabbing}
Le \= x \= x \= x \= x \= x \= x \= x \= x \= \kill
\qe($x:$ \textbf{term}~; $A:$ \textbf{set of terms}); \\
1. \> $R$ $\leftarrow$ $I(x)$ \\
2. \> \textbf{for each} hyperedge $\langle \{u_1,...,u_r\},x\rangle$ in $\ACCA$
\textbf{do} \\
3. \> \> \textbf{if} $\{u_1,...,u_r\}\cap A=\emptyset$ \textbf{then} $R$
$\leftarrow$ $R$ $\cup$ (\qe$(u_1,A\cup\{u_1\})$
$\cap$ $\ldots$ $\cap$ \qe$(u_r,A\cup\{u_r\}))$ \\
4. \> \textbf{return}($R$)
\end{tabbing}
}
\caption{The procedure \qe} \label{fig:esmmt}
\end{figure}

The procedure \qe, presented in Figure~\ref{fig:esmmt}, computes $\ans(t,S)$ for
a given term $t$ (and an implicitly given source $S$) by applying in a
straightforward way Corollary~\ref{prop:iteli}.  To this end, \qe~must be
invoked as \qe$(t,\{t\}).$ The second input parameter of \qe~is the set of terms
on the \emph{path} from $t$ to the currently considered term $x.$ This set is
used to guarantee that $t$ is connected to all terms considered in the recursion
by a cycle-free simple path. \qe~accumulates in $R$ the result. The
correctness of \qe~ can be established by just observing that, for all objects
$o\in\obj,$ $o$ is in the set $R$ returned by \qe$(t,\{t\})$ if and only if $o$
satisfies the two conditions expressed by Corollary~\ref{prop:iteli}.

As an example, let us consider the sequence of calls made by the procedure \qe~in
evaluating the query $a2$ in the example source, as shown in
Table~\ref{tab:esmmt}.  The calls marked with a $\star$ are those in which the
test in line 3 gives a negative result.  Upon evaluating \qe$(c2,\{a2,b1,c2\})$
the procedure realizes that the only incoming hyperedge in $c2$ is $\langle
\{b1,b3\},c2\rangle,$ whose tail $\{b1,b3\}$ has a non-empty intersection with
the current path $\{a2,b1,c2\};$ so the hyperedge is ignored.  In this case, the
cycle $(b1,c2,b1)$ is detected and properly handled.  Analogously, upon
evaluating \qe$(b1,\{a2,b2,c2,b1\}),$ the cycle $(c2,b1,c2)$ is detected and
properly handled. Also notice the difference between the calls
\qe$(c2,\{a2,b1,c2\})$ and \qe$(c2,\{a2,b2,c2\}).$ The both concern $c2,$ but in
the former case, $c2$ is encountered upon descending along the path $(a2,b1,c2)$
whose next hyperedge is $\langle \{b1,b3\},c2\rangle;$ following that hyperedge,
would lead the computation back to the node $b1,$ which has already been met,
thus the result of the call is just $I(c2).$ In the latter case, $c2$ is
encountered upon descending along the path $(a2,b2,c2),$ thus the hyperedge
leading to $b1$ and $b3$ must be followed, since none of the terms in its tail
have been touched upon so far.

\begin{center}
\begin{table}
\caption{Evaluation of \qe$(a2,\{a2\})$ \label{tab:esmmt}}
\begin{tabular}{||r|l||} \hline\hline
Call & Result \\
\hline
\qe$(a2,\{a2\})$ & $I(a2)~\cup$ \qe$(b3,\{a2,b3\}) ~\cup$
(\qe$(b1,\{a2,b1\}) ~\cap$ \qe$(b2,\{a2,b2\}))$ \\
\qe$(b3,\{a2,b3\})$ & $I(b3)$ \\
\qe$(b1,\{a2,b1\})$ & $I(b1) ~\cup$ \qe$(c1,\{a2,b1,c1\}) ~\cup $
\qe$(c2,\{a2,b1,c2\})$ \\
\qe$(b2,\{a2,b2\})$ & $I(b2) ~\cup$ (\qe$(c2,\{a2,b2,c2\}) ~\cap$
\qe$(c3,\{a2,b2,c3\}))$ \\
\qe$(c1,\{a2,b1,c1\})$ & $I(c1)$ \\
\qe$(c2,\{a2,b1,c2\})$ & $I(c2)$ $\star$ \\
\qe$(c2,\{a2,b2,c2\})$ & $I(c2) ~\cup$ (\qe$(b1,\{a2,b2,c2,b1\}) ~\cap$
\qe$(b3,\{a2,b2,c2,b3\}))$ \\
\qe$(c3,\{a2,b2,c3\}))$ & $I(c3)$ \\
\qe$(b1,\{a2,b2,c2,b1\})$ & $I(b1) ~\cup$ \qe$(c1,\{a2,b2,c2,b1,c1\})\star$ \\
\qe$(b3,\{a2,b2,c2,b3\}))$ & $I(b3)$ \\
\qe$(c1,\{a2,b2,c2,b1,c1\})$ & $I(c1)$ \\
\hline\hline
\end{tabular}
\end{table}
\end{center}

From a complexity point of view, \qe~visits all terms that lie on a cycle-free
simple path ending at the query term $t$ in the taxonomy B-graph $\ACCA.$ As
shown in Section \ref{sec:answerSet}, the number of such terms can be exponential in the
size of the taxonomy. For each term, \qe~performs set-theoretic operations on
sets of objects, which have polynomial time complexity. Thus, though \qe~ operates in exponential time in the size of the taxonomy, 
it has  polynomial time complexity in the size of the information source.

From a more practical point of view, there is an obvious alternative to \qe~for
computing $\ans(t,S),$ that is to solve the decision problem for each object
$o\in\obj.$ However, this method is not practically applicable to peer-to-peer
networks, thus we do not take it into consideration any longer.

\subsection{Negation}
\label{cha:neg}

In this section we deal with negation. We first consider the addition of
negation to the taxonomy of the source, then the simpler case in which negation
is used in queries only.

\subsubsection{Adding negation to the taxonomy}
\label{sec:negtax}

If the queries in taxonomy relationships have negation, then the source
corresponds to a datalog program with rules that contain negation in their
bodies, and it is well known (e.g. see~\cite{Ullman88I}) that such programs may
not have a unique minimal model.  This is illustrated by the source shown in
Figure \ref{fig:MModels}: the left part shows the source taxonomy, while the
right part shows the source interpretation, $I,$ and two minimal models $I_a$ and
$I_b.$

\begin{figure}[htbp]
  \centering
  \begin{minipage}{5cm}
    \begin{picture}(4,2)
      \put(0.25,1.25){\makebox(1.5,0.5){$a2\wedge\neg a1$}}
      \put(1.75,1.5){\vector(1,0){0.5}}
      \put(0.35,0.25){\makebox(0.5,0.5){$a2$}}
      \put(1.25,0.25){\makebox(0.5,0.5){$a1$}}
      \put(2.25,0.25){\makebox(1.5,0.5){$b2\wedge\neg b1$}}
      \put(2.25,0.5){\vector(-1,0){0.5}}
      \put(2.25,1.25){\makebox(0.5,0.5){$b1$}}
      \put(3.25,1.25){\makebox(0.5,0.5){$b2$}}
    \end{picture}
  \end{minipage}
  \begin{minipage}{6cm}
    \begin{tabular}{||c|c||c|c||} \hline\hline
      query&    $I$           & $I_a$     & $I_b$ \\\hline
      $a1$      & $\emptyset $  & $\{1\}$   & $\emptyset $ \\
      $a2$      & $\{1\}$       & $\{1\}$   & $\{1\}$    \\
      $b1$      & $\emptyset $  & $\emptyset$& $\{1\}$ \\
      $b2$      & $\{1\}$       & $\{1\}$   & $\{1\}$    \\
    $b2 \wedge\neg b1$ & $\{1\}$       & $\{1\}$   & $\emptyset $ \\
    $a2 \wedge\neg a1$ & $\{1\}$       & $\emptyset$& $\{1\}$    \\\hline\hline
    \end{tabular}
  \end{minipage}
\caption{A source with no unique minimal model}
  \label{fig:MModels}
\end{figure}
The lack of a unique minimal model turns out to be a serious drawback. Let
$\EL_T^\neg$ be the language of conjunctive queries in which negations of terms
may occur, \ie~$\EL_T^\neg$ is given by (as usual, $t$ is a term in $T$):
\begin{tabbing}
  ind \= m \= ::= \= \kill
  \> $q$ \> ::= \> $d ~ | ~ q \vee d$ ~~\= ($q$ is a \emph{query}) \\
  \> $d$ \> ::= \> $l ~ | ~ l \wedge d$ \> ($d$ is a \emph{disjunct}) \\
  \> $l$ \> ::= \> $t ~ | ~\neg t$ \> ($l$ is a \emph{literal}).
\end{tabbing}
Moreover, let $\C_T^\neg$ be the sub-language of $\EL_T^\neg$ consisting of just
disjuncts. A \emph{neg-extended taxonomy} is a pair $(T, \preceq^\neg)$, where
$T$ is a terminology and $\preceq^\neg \subseteq (\C_T^\neg\times \C_T^\neg)$
is reflexive and transitive, such that if $q_1\preceq^\neg q_2$ and $q_1\not= q_2,$
then $q_2=t$ for some term $t\in T.$ A \emph{neg-extended source} $S$ is a
4-tuple $(T,\preceq^\neg,\obj,I),$ where $(T,\preceq^\neg)$ is a neg-extended
taxonomy and $I$ is an interpretation for it.

It can be proved that:
\begin{proposition}\label{prop:NPhard}
  \emph{Deciding whether an object $o\in \obj$ is in the answer of a query
    $q\in\EL_T^\neg$ in a neg-extended source $S,$ $o\in \ans(q,S),$ is a
    coNP-hard problem.}
\end{proposition}
The proof is based on the following polynomial reduction from SAT. Let $\alpha$
be a CNF formula of propositional logic over an alphabet $V,$ that is:
\[
\alpha = \bigwedge_{i=1}^n\alpha_i \hspace*{1cm}\alpha_i = \bigvee_{j=1}^{m_i}
l_{ij}
\]
where $l_{ij}$ is either a positive literal, that is a letter $v\in V,$ or a
negative literal, that is $\neg u$ where $u\in V.$ We map $\alpha$ into a
neg-extended source $S_\alpha=(T_\alpha,\preceq_\alpha,\obj_\alpha,I_\alpha),$
and a query $q_\alpha$ as follows:
\begin{itemize}
\item $T_\alpha = V;$
\item $\obj_\alpha=\{1\};$
\item the query $q_\alpha$ is given by
\[
\bigvee\{v_1\wedge\ldots\wedge v_k~|~ \neg v_1\vee\ldots\vee\neg v_k \mbox{ is a
  conjunct $\alpha_i$ in }\alpha \; (v_i \in V) \}.
\]
If there is no conjunct $\neg v_1\vee\ldots\vee\neg v_k$ in $\alpha,$ then
let $\alpha_1$ be $l_1\vee \ldots\vee l_k;$ we then set 
$q_\alpha =\overline{l_1} \wedge\ldots\wedge \overline{l_k},$ where $\overline{\neg u}=u$
and $\overline{v}=\neg v;$
\item for each remaining conjunct $\alpha_i$ in $\alpha,$
\begin{enumerate}
\item if $\alpha_i$ is a letter $v,$ then $I_\alpha(v)=\{1\};$ if for no
  conjunct $\alpha_i,$ $\alpha_i=v,$ then $I_\alpha(v)=\emptyset;$
\item if $\alpha_i$ is $l_1\vee\ldots\vee l_k$ for $k\geq 2,$ where at least one
  literal is positive, say w.l.o.g. that $l_1$ is the positive literal $u,$ then
  the subsumption relationship
  $(\overline{l_2}\wedge\ldots\wedge\overline{l_k},u)$ is in $\preceq^r_\alpha.$
\end{enumerate}
\end{itemize}
For instance, the propositional formula
\begin{eqnarray*}
\alpha & = & a2 \wedge b2 \wedge \\
& & (a1 \vee \neg a2 \vee b1) \wedge (a1 \vee b1 \vee \neg b2) \wedge \\
& & \neg a1 \wedge \neg b1
\end{eqnarray*}
is mapped into the source shown in Figure~\ref{fig:MModels} and the query
$q_\alpha=a1\vee b1.$ We now show the following

\medskip\noindent\textbf{Lemma} $1\in \ans(q_\alpha,S_\alpha)$ iff $\alpha$ is
unsatisfiable. \\
In fact, we prove the equivalent form: $1\not\in
\ans(q_\alpha,S_\alpha)$ iff $\alpha$ is satisfiable. \\
($\rightarrow$) Suppose $\alpha$ is satisfiable, and let $f$ be a truth
assignment over $V$ satisfying it. Let $J$ be the interpretation of the
taxonomy $(T_\alpha,\preceq_\alpha)$ such that, for each term $t\in V,$
\[
J(t)=\left\{
  \begin{array}{ll}
\{1\}     & \mbox{if $f(t)=T$} \\
\emptyset & otherwise
  \end{array}
\right.
\]
We have that $I_\alpha\leq J,$ since for each $t\in V,$ either $I_\alpha(t)$ is
empty, or $I_\alpha(t)=\{1\}.$ In the former case, $I_\alpha(t)\subseteq J(t)$
for any $J(t).$ In the latter case, we have that $\alpha_j=t$ for some $1\leq
j\leq n,$ which implies $f(t)=T$ (since $f$ satisfies $\alpha$) which implies
$J(t)=\{1\}$ and again $I_\alpha(t)\subseteq J(t).$ Moreover,
$(q,u) \in \preceq_\alpha$ implies $J(q)\subseteq J(u).$ In proof,
$(q,u) \in \preceq_\alpha$ iff $\alpha_k=\neg q\vee u$ for some $1\leq k\leq n,$
which implies $f(\neg q\vee u)=T$ (since $f$ satisfies $\alpha$) and therefore:
either $f(\neg q)=T$ and by construction $J(q)=\emptyset,$ or $f(u)=T$ and by
construction $J(u)=\{1\};$ in both cases $J(q)\subseteq J(u).$ Hence $J$ is a
model of $S_\alpha.$ However, $1\not\in J(q_\alpha).$ In fact, by construction,
for any disjunct $d$ in $q_\alpha,$ there exists $\alpha_j=\neg d$ for some
$1\leq j\leq n.$ Since $f$ satisfies $\alpha,$ it follows that $f$ satisfies
$\neg d$ so $f(d)=F.$ But then $J(d)=\emptyset$ for each disjunct $d$ in
$q_\alpha,$ which implies $J(q_\alpha)=\emptyset.$ So, $1\not\in J(q)$ for a
model $J,$ that is $1\not\in \ans(q_\alpha,S_\alpha).$ \\
($\leftarrow$) Suppose $1\not\in \ans(q_\alpha,S_\alpha),$ and let $J$ be a
model of $S_\alpha$ such that $1\not\in J(q_\alpha).$ Let $f$ be the truth
assignment over $V$ defined as follows, for each letter $t\in V,$
\[
f(t)=\left\{
  \begin{array}{ll}
T & \mbox{if $1\in J(t)$} \\
F & otherwise
  \end{array}
\right.
\]
By a similar argument to the one developed in the \emph{if} part of the proof,
it can be proved that $f$ satisfies $\alpha,$ and this
completes the proof of the Lemma.

From the last Lemma and the NP-completeness of SAT, the coNP-hardness of
deciding query re-writing in neg-extended sources follows. \ok

We observe that it is essential for the reduction that the query language allows
negation. Otherwise, propositional formulae which do not have a conjunct
consisting of all negative literals, such as $\neg v_1\vee\ldots\vee\neg v_k,$
could not be reduced.

\subsubsection{Adding negation in queries}
\label{sec:t2tneg}

In this Section, we consider the evaluation of queries containing negation over
a source. To this end, we need first to define the extension of a negative
literal in an interpretation $I.$ The obvious way of doing so is as follows:
$I(\neg t)=\obj\setminus I(t).$ However, as it is well-known, if we maintain our
definition of query answer, as $\ans(q,S)=\{o\in \obj~|~o\in J(q) \mbox{ for all
  models $J$ of $S$}\},$ a negative literal in a query is equivalent to the
\emph{false} clause, because there is not enough information in the taxonomy of
a source to support a negative fact.

In order to derive an intuitive and, at the same time, logically well-grounded
evaluation procedure for extended queries, we need an alternative query
semantics (\ie~$\ans$). In order to define it, let us consider a logical
reformulation of the problem in terms of datalog. We map each term $t_i$ into
two predicate symbols:
\begin{itemize}
\item an extensional one, denoted $\mathtt{C_{t_i}},$ representing the
  interpretation of $t_i,$ \ie~$I(t_{t_i});$ and
\item an intensional one, denoted $\mathtt{Y_{t_i}},$ representing $t_i$ in the
  rules encoding the subsumption relation.
\end{itemize}
The obvious connection between $\mathtt{C_{t_i}}$ and $\mathtt{Y_{t_i}}$ is that all
facts expressed via the former are also true of the latter, and this is captured
by stating a rule (named ``extensional'' below) of the form
$\mathtt{C_{t_i}}(x)\rightarrow\mathtt{Y_{t_i}}(x)$ for each term $t_i.$

\begin{definition}[Source program]
  \emph{Given a source $S=(T,\preceq,\obj,I),$ the \emph{source program}
    of $S$ is the set of clauses $P_S$ given by $P_S=TR_S\cup ER_S\cup F_S,$
    where:
  \begin{itemize}
  \item $TR_S=\{\mathtt{Y_{t}(x) :-~ Y_{t_1}(x)},\ldots,\mathtt{Y_{t_m}(x)}~|~t_1\wedge
    \ldots\wedge t_m\preceq^r t\}$ are the \emph{terminological rules} of $P_S;$
  \item $ER_S=\{\mathtt{Y_{t_i}(x) :-~ C_{t_i}(x)}~|~t_i\in T\}$ are the
    \emph{extensional rules} of $P_S;$
  \item $F_S=\{\mathtt{C_{t_i}(o)}~|~o\in I(t_i)\}$ are the \emph{facts} of $P_S,$
    stated in terms of constants \texttt{o} which are one-to-one with the
    elements of $\obj$ (unique name assumption). \ok
  \end{itemize}
}
\end{definition}

Next, we translate queries in the language $\EL_T.$

\begin{definition}[Query program] \label{def:QueryProg}
  \emph{Given a query $q\in \EL_T$ to a simple source $S=(T,\preceq,\obj,I),$
    the \emph{query program} of $q$ is the set of clauses $P_q$ given by:
\[
\{\mathtt{q(x):-~Y_{t_1}(x),\ldots,Y_{t_k}(x)}~|~t_1\wedge\ldots\wedge t_k \mbox{ is
  a disjunct of } q\}.
\]
where \texttt{q} is a new predicate symbol.} \ok
\end{definition}

In order to show the equivalence of the original model with its datalog
translation, we state the following:

\begin{proposition} \label{prop:equivalence}
  \emph{For each source $S=(T,\preceq,\obj,I),$ and query $q\in \EL_T,$
    $\ans(q,S) =\{o\in \obj~|~P_S\cup P_q\models \mathtt{q(o)}\}.$ \ok }
\end{proposition}

Let us consider this mapping in light of the new query language $\EL_T^\neg.$
The source program $P_S$ remains a pure datalog program, while the query program
$P_q$ of any query $q \in \EL_T^\neg$ against $S$ becomes:
\[
\{\mathtt{q(x):-~L_{v_1}(x),\ldots,L_{v_k}(x)}~|~v_1\wedge\ldots\wedge v_k \mbox{ is
  a disjunct of } q\}
\]
where each $\mathtt{L_{v_i}}$ is either $\mathtt{Y_{v_i}}$, if $v_i=t_i$, or
$\mathtt{\neg Y_{v_i}}$, if $v_i=\neg t_i$ ($t_i \in T$).

This kind of queries are dealt with by using an approximation of CWA, which
can be characterized either procedurally, in terms of program stratification, or
declaratively, in terms of perfect model. We will adopt the former
characterization.  In fact, $P_q$ is a datalog$^\neg$ program, and so is the
program $P_S\cup P_q.$ The latter program is stratified, by the level mapping
$l$ defined as follows:
\[
l(pred)=\left\{
  \begin{array}{ll}
1 & \mbox{if \emph{pred} is }\mathtt{q} \\
0 & \mbox{otherwise}
  \end{array}
\right.
\]
It follows that $P_S\cup P_q$ has a minimal Herbrand model $M_S^q$ given by
(\cite{gct1990}) the least fixpoint of the transformation $T^\prime_{P_q\cup
  M_{P_S}}$ where $M_{P_S}$ is the least Herbrand model of the datalog program
$P_S,$ and $T^\prime_P$ is the extension to datalog$^\neg$ of the $T_P$
operator, on which the standard semantics of pure datalog is based. The model
$M_S^q$ is found from $M_{P_S}$ in one iteration since only instances of
\texttt{q} are added at each iteration, and \texttt{q} does not occur in the
body of any rule.  The following definition establishes an alternative notion of
answer for queries including negation.

\begin{definition}[Extended answer]
  \emph{Given an extended query $q$ to a source $S=(T,\preceq,\obj,I),$ the
    \emph{extended answer to $q$ in} S, denoted $\varepsilon(q,S),$ is given by:
    $\varepsilon(q,S) =\{o\in \obj~|~M_S^q\models \mathtt{q(o)}\}$ \ok }
\end{definition}

We conclude by showing how extended answers can be computed.

\begin{proposition} \label{prop:ExtAnsComp}
  \emph{For each source $S=(T,\preceq,\obj,I),$ and query $q\in
    \EL_T^\neg,$ $\varepsilon(q,S)$ is given by:
  \begin{enumerate}
  \item $\varepsilon(q\vee d,S)=\varepsilon(q,S)\cup \varepsilon(d,S),$
  \item $\varepsilon(l\wedge d,S)=\varepsilon(l,S)\cap \varepsilon(d,S),$
  \item $\varepsilon(t,S)=\bim(t),$
  \item $\varepsilon(\neg t,S)=\obj\setminus \varepsilon(t,S).$ \ok
  \end{enumerate}
}
\end{proposition}

From a practical point of view, computing $\varepsilon(\neg
t_1\wedge\ldots\wedge\neg t_k)$ requires computing:
\[
\obj\setminus (\bim(t_1)\cup\ldots\cup\bim(t_k))
\]
which in turn requires knowing $\obj,$ \ie~the whole set of objects of the
network. As this knowledge may not be available, or may be too expensive to
obtain, one may want to resort to a query language making a restricted usage of
negation, for instance by forcing each query disjunct to contain at least one
positive term.

\subsection{Disjunctive information sources}
\label{sec:q2qdl}

In this section we consider disjunctive sources, whose taxonomies allow
subsumption relationships between queries.  Formally, a \emph{disjunctive
  taxonomy} is a pair $(T, \preceq_d)$ where $T$ is a terminology and
$\preceq_d\subseteq (\EL_T\times \EL_T)$ is reflexive and transitive. A
\emph{disjunctive source} $S$ is a 4-tuple $(T,\pd,\obj,I)$ where
$(T,\pd)$ is a disjunctive taxonomy and $(\obj,I)$ is an interpretation for it.

Disjunctive sources may not have a unique minimal model. As an example, the
source $(T,\pd,\obj,I)$ where:
\begin{itemize}
\item $T=\{a1,a2,b1,b2\}$
\item $\pd^r=\{(a2,a1\vee b1),(b2,a1\vee b1)\}$
\item $\obj=\{1\}$ and
\item $I=\{(a2,\{1\}),(b2,\{1\})\}$
\end{itemize}
has two minimal models, $I_1=I\cup\{(a1,\{1\})\}$ and $I_2=I\cup\{(b1,\{1\})\}.$

Loosing the uniqueness of the minimal model is enough to make query evaluation
for this kind of sources computationally difficult.

\begin{proposition}\label{prop:NPhard2}
  \emph{Deciding whether an object $o\in \obj$ is in the answer of a query
    $q\in\EL_T$ in a disjunctive source $S,$ $o\in \ans(q,S),$ is a coNP-hard
    problem.}
\end{proposition}
The proof is similar to that of Proposition \ref{prop:NPhard}. For brevity, we just
show the reduction from SAT. Let $\alpha$ be as in the proof of
Proposition~\ref{prop:NPhard}. We map $\alpha$ into a
disjunctive source $S_\alpha=(T_\alpha,\preceq_\alpha,\obj_\alpha,I_\alpha),$
and a query $q_\alpha$ as follows:
\begin{itemize}
\item $T_\alpha = V;$
\item $\obj_\alpha=\{1\};$
\item the query $q_\alpha$ is given by
  \begin{center}
$\bigvee\{v_1\wedge\ldots\wedge v_k| \neg v_1\vee\ldots\vee\neg v_k \mbox{ is a
  conjunct in }\alpha \; (v_i \in V)\} ~\vee$ \\ $\bigvee \{\neg u_1\wedge\ldots\wedge \neg u_k~|~
  u_1\vee\ldots\vee u_k \mbox{ is a conjunct in }\alpha \; (u_i \in V)\}$
  \end{center}
If there are no such conjuncts $\neg v_1\vee\ldots\vee\neg v_k$ or $\neg
u_1\wedge\ldots\wedge \neg u_k$ in $\alpha,$ then let $\alpha_1$ be $l_1\vee
\ldots\vee l_k;$ we then set $q_\alpha = \overline{l_1} \wedge\ldots\wedge
\overline{l_k},$ where $\overline{\neg u}=u$ and $\overline{v}=\neg v.$
\item for each remaining conjunct $\alpha_i$ in $\alpha,$
\begin{enumerate}
\item if $\alpha_i$ is a letter $v,$ then $I_\alpha(v)=\{1\};$ if for no
  conjunct $\alpha_i,$ $\alpha_i=v,$ then $I_\alpha(v)=\emptyset;$
\item if $\alpha_i$ is $\neg u_1\vee\ldots\vee \neg u_j\vee v_1\vee\ldots\vee
  v_m$ where $j,m\geq 1$ then the subsumption relationship
  $(u_1\wedge\ldots\wedge u_j,v_1\vee\ldots\vee v_m)$ is in $\preceq^r_\alpha.$
\end{enumerate}
\end{itemize}
In the present case, the propositional formula
\begin{eqnarray*}
\alpha & = & a2 \wedge b2 \wedge \\
& & (a1 \vee \neg a2 \vee b1) \wedge (a1 \vee b1 \vee \neg b2) \wedge \\
& & \neg a1 \wedge \neg b1
\end{eqnarray*}
is mapped into the source shown in the previous example.

It can be shown that $1\in \ans(q_\alpha,S_\alpha)$ iff $\alpha$ is
unsatisfiable.

\section{Networks of Information Sources}
\label{sec:Mediators}

In this Section we introduce networks of information sources. The model is first
outlined, and then query evaluation is considered.

\subsection{The model}
\label{sec:netmodel}

In order to be a component of a networked information system, a source is
endowed with additional subsumption relations, called articulations, which
relate the source terminology to the terminologies of other sources of the same
kind.

\begin{definition}[Articulation] \label{def:Articulation}
  \emph{Given two terminologies $T$ and $U,$ an \emph{articulation} from $T$ to
    $U,$ $\preceq_{TU},$ is a non-empty binary relation from $\EL_{U}$ to $T,$
    such that $q \preceq_{TU} t$ implies that $q$ is a conjunctive query.  } \ok
\end{definition}

An articulation relationship is not syntactically different from a subsumption
relationship, except that its head may be a term of a different terminology than
the one where the terms making up its tail come from.

\begin{definition}[Articulated Source] \label{def:ArticulatedSource}
  \emph{An {\em articulated source} $\ES$ over $k\geq 0$ disjoint terminologies
    $T_1,...,T_k,$ is a 5-tuple $\ES=(T_\ES,\preceq_\ES,\obj,I_\ES,R_\ES),$
    where:
    \begin{itemize}
    \item $(T_\ES, \preceq_\ES,\obj,I_\ES)$ is a source;
    \item $R_\ES$ is a set of articulations $R_\ES=\{\preceq_{T_\ES,T_1},\ldots,
      \preceq_{T_\ES,T_k}\}.$ \ok
    \end{itemize}
  }
\end{definition}

Articulations are used to connect an articulated source to other articulated
sources, so creating a networked information system. An articulated source $\ES$
with an empty stored interpretation, \ie~ $I_\ES(t)=\emptyset$ for all $t\in
T_\ES,$ is called a \emph{mediator} in the literature.

\begin{definition}[Network] \label{def:Network}
  \emph{A \emph{network of articulated sources,} or simply a \emph{network},
    $\N$ is a non-empty set of articulated sources $\N=\{\ES_1,\ldots,\ES_n\},$
    where each $\ES_i$ is articulated over the terminologies of some of the
    other sources in $\N$ and all terminologies $T_{\ES_1},\ldots,T_{\ES_n}$ of
    the sources in $\N$ are disjoint.}  \ok
\end{definition}

Notice that the domain of the interpretation of an articulated source is
independent from the source, thus the same for any articulated source. This is
not necessary for our model to work, just reflects a typical situation of
networked resources such as URLs. Relaxing this constrain would have no impact
on the results reported in the present study.

Since in a network: (a) there is no source acting at the global level, (b) all
sources store data, and (c) as we will see, data are exchanged via direct
communication, each source can be seen as, and will in fact be called, a
\emph{peer,} and the network as a \emph{peer-to-peer} information system.
Articulations of the network peers will also be referred as {\em P2P mappings}.

An intuitive way of interpreting a network is to view it as a single source
which is distributed along the nodes of a network, each node dealing with a
specific vocabulary. The global source can be logically constructed by removing
the barriers which separate local sources, as if (virtually) collecting all the
network information in a single repository.  The notion of \emph{network source}
captures this interpretation of a network.

\begin{definition}[Network source] \label{def:NetworkSource}
  \emph{The \emph{network source} $S_\N$ of a network of articulated sources
    $\N=\{\ES_1,\ldots,\ES_n\},$ is the source \\
    $S_\N=(T_\N,\sqsubseteq,\obj,I_\N),$ where:
    \begin{itemize}
    \item $T_\N = \bigcup_{i=1}^n T_{\ES_i};$
    \item $I_\N  =  \bigcup_{i=1}^n I_{\ES_i}$
    \item $\sqsubseteq  = ( \bigcup_{i=1}^n \sqsubseteq_{\ES_i} )^*$
    \end{itemize}
    where $\sqsubseteq_{\ES_i}$ is the {\em total subsumption} of the source $\ES_i,$
    given by the union of the subsumption relation $\preceq_{\ES_i}$ with all
    articulations of the source, that is:
    \[
    \sqsubseteq_{\ES_i} \;= \; \preceq_{\ES_i} \; \cup \; \bigcup R_{\ES_i}
    \]
    and $A^*$ denotes the transitive closure of the binary relation $A.$ A
    \emph{network query} is a query over $T_\N.$ } \ok
\end{definition}

It is not difficult to see that $\s$ is reflexive and transitive, and every
non-trivial subsumption relationship in it relates a conjunctive query in anyone
of the terminologies $T_{\ES_1},\ldots,T_{\ES_n}$ to a single term. Thus, $S_\N$ is indeed a
source. Such source emerges in a bottom-up manner from the articulations of the
peers.  This distinguishes peer-to-peer systems from federated distributed
databases.

A \emph{network query} $q$ is a query in anyone of the query languages supported
by the network, that is $q\in\EL_{T_{\ES_i}}$ for some $i\in[1,n].$ As it will be
evident, the method that we will set up only requires minor modifications to be
able to evaluate also queries in the language $\EL_{T_\N},$ that is queries that
mix terms from different terminologies. We do not provide this facility because
it does not seem to make much sense in our vision.

The answer to a network query $q,$ or \emph{network answer,} is given by
$\ans(q,S_\N).$

\begin{figure}[htbp]
  \centering
  \begin{picture}(6.5,3)
    \put(0.75,2.25){\makebox(0.5,0.5){$a1$}}
    \put(1.25,1.25){\makebox(0.5,0.5){$a2$}}
    \put(0.25,1.25){\makebox(0.5,0.5){$a3$}}
    \put(2.5,2.25){\makebox(1,0.5){$b1\wedge b2$}}
    \put(3.5,1.25){\makebox(1,0.5){$b1\wedge b3$}}
    \put(2.5,1.25){\makebox(0.5,0.5){$b3$}}
    \put(4,2.25){\makebox(0.5,0.5){$b1$}}
    \put(3.75,0.25){\makebox(0.5,0.5){$b2$}}
    \put(5.25,2.25){\makebox(0.5,0.5){$c1$}}
    \put(5.25,1.25){\makebox(0.5,0.5){$c2$}}
    \put(5.25,0.25){\makebox(1,0.5){$c2\wedge c3$}}

    \put(0.5,1.75){\vector(1,2){0.25}}
    \put(1.5,1.75){\vector(-1,2){0.25}}
    \put(2.5,2.25){\vector(-3,-2){0.75}}
    \put(2.5,1.5){\vector(-1,0){0.75}}
    \put(5.2,2.5){\vector(-1,0){0.75}}
    \put(5.25,1.65){\vector(-1,1){0.75}}
    \put(4.55,1.5){\vector(1,0){0.7}}
    \put(5.1,0.5){\vector(-1,0){0.85}}
    \put(1,2){\oval(2,2){}}
    \put(3.5,1.5){\oval(2.5,3){}}
    \put(5.75,1.5){\oval(1.5,3){}}
    \put(0,0.65){\makebox(0.5,0.35){$P_a$}}
    \put(1.9,0){\makebox(0.5,0.35){$P_b$}}
    \put(6.35,0){\makebox(0.5,0.35){$P_c$}}
  \end{picture}
  \caption{A network taxonomy}
  \label{fig:net}
\end{figure}
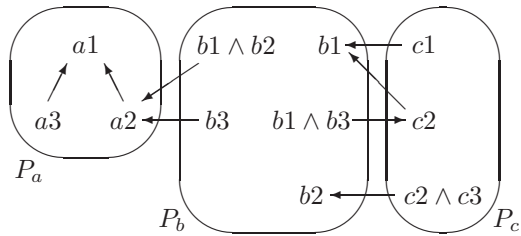

Figure~\ref{fig:net} presents the taxonomy of a network source $S_\N,$ where
$\N$ consists of 3 peers $\N=\{P_a,P_b,P_c\}.$ As it can be verified, this is
the same taxonomy as the one shown in Figure~\ref{fig:extsource}, except that
now some of its subsumption relationships are elements of articulations.

\subsection{Network query evaluation}
\label{sec:nqeval}

This Section presents a network query evaluation procedure based on the method
devised in the centralized case. First, a functional model of each peer is
introduced, then the algorithms corresponding to the operations on the interface
of the peer are given. Correctness and complexity of these algorithms are
discussed in Section~\ref{sec:cc}, while Section~\ref{sec:opt} concludes by
considering optimization issues.

\subsubsection{The functional model of a peer}
\label{sec:func}

In order to illustrate our query evaluation procedure, we now define a peer from
a functional point of view. In this respect, we see a peer as a software
component uniquely identified in the network by a peer ID.  The interface of a
peer exposes just one method:
\begin{itemize}
\item \query, which takes as input a network query $q$ and evaluates it,
  returning the set of objects $\ans(q,S_\N).$
\end{itemize}
The user (whether human or application program) is supposed to use this method
for the evaluation of network queries. We assume that $q$ is expressed in the
query language of the peer. As it will be argued in due course, this assumption
can be relaxed without any substantial change to our framework.

In addition to \query, a peer has methods for sending to or receiving messages
from other peers. We do not enter into the details of these methods: there are
several options, which do not make any difference from the point of view of our
model. Instead, we detail the types of messages that can be exchanged between
peers. These can be of one of the following 2 types:
\begin{itemize}
\item \ask: by sending a message of this kind to a peer $P,$ the present peer
  asks $P$ to evaluate a term query on $P$'s query language. The receiving peer
  $P$ processes \ask~messages according to the \qe~procedure (Figure \ref{fig:esmmt}), as we will see in
  detail below. An \ask~message has the following fields:
  \begin{itemize}
  \item \pid: the id of the present peer, which is sending the message;
  \item \qid: the id of the query that \pid~ is sending for evaluation;
  \item \emph{t:} the query term of \qid;
  \item \emph{A:} the set of already visited terms. These two last parameters
    are those of the \qe~procedure.
  \end{itemize}
\item \tell: by sending a message of this kind to a peer $P,$ the present peer
  returns to $P$ the result of the evaluation of a term query which had
  previously been \ask-ed by $P.$ A \tell~message has the following fields:
  \begin{itemize}
  \item \qid: the ID of the query whose result is being returned;
  \item \emph{RES:} the set of objects resulting from the evaluation of \qid.
  \end{itemize}
\end{itemize}

We will denote the sending of a message of one of these two kinds $m$ to the
peer $P$ as $P$\texttt{:}\emph{m(field values)}.  By decoupling the request of
evaluation from the return of the result, we aim at minimizing the number of
sessions open at any time between peers, thus removing a serious obstacle
towards scalability.  \query~ does not follow this paradigm since it involves
only a local interaction.

Each peer processes the incoming messages depending on their type and content.
In order to carry out this work, the peer keeps a \emph{(query) log,} that is a set of
objects, each associated to a query in whose evaluation the peer is currently
involved. A log object has the following attributes:
\begin{itemize}
\item \pid: the id of the peer who sent the query (can be the local peer itself);
\item \qid: the id of the query;
\item \emph{t:} the query term (we recall that we need to deal only with term
  queries);
\item \emph{n:} the number of open calls in \qid~ (see next paragraph);
\item \qp: the query program representing the current status of evaluation of
  \qid.  A query program is a set of \emph{sub-programs}
  $\{\subp_1,\ldots,\subp_k\}$ where each sub-program $\subp_j$ is a set of
  \emph{calls.} A call is a sub-query of \qid, and can be:
  \begin{itemize}
  \item \emph{open,} meaning that the sub-query is being evaluated, in which
    case the call is the sub-query id; or
  \item \emph{closed,} meaning the sub-query has been evaluated, in which case
    the call is the resulting set of objects.
  \end{itemize}
\end{itemize}
Since no two log objects can have the same query id, we will represent a log
object as a 5-tuple (\pid,\qid,\emph{t},\emph{n},\qp).

\subsubsection{\query}
\label{sec:query}

Let us assume that the input query $q$ posed to a peer $\ES$, is given by
\[
q = \bigvee C_i
\]
where each $C_i$ is a conjunctive query. As a first step, \query~reduces $q$ to
a term query $t$ by generating a new term $t$ not in $T_\N$ and
inserting a new hyperedge $(C_i, t)$ into the local taxonomy B-graph
(i.e. that corresponding to $(T_{\ES},  \sqsubseteq_{\ES})$), for each
conjunctive query $C_i$ in $q.$ This work is carried out by the function
\textsc{Modify-taxonomy,} which returns the newly generated term $t.$ A new
query id for $t$ is subsequently obtained by \query, and an \ask~ message is
sent to the peer itself for evaluating $t.$ As required by \qe, the set of
already visited terms consists just of $t$ itself. At this point \query~hangs on
the log, until the log object associated to the query $t$ is closed, that is the
number of its open call is 0. Notice that this object is created only after the
\ask~message sent on line 3 is processed, but this creates no problem, as all
\query~has to do in the meantime is wait. When the log object is finally closed,
\query~retrieves it and deletes it from the log, by using the function
\textsc{Delete,} which returns the object itself. When the object is closed, its
query program, that is the value of the last field, equals to $\ans(t,S_\N).$
This value is assigned to the variable $R.$ On line 6, the subsumption
relationships inserted by \textsc{Modify-taxonomy} are removed by
\textsc{Cleanup-taxonomy,} and $R$ is finally returned.

\begin{figure}[htbp]
{\small
  \begin{tabbing}
    Les \= xxxx \= xxxx \= xxxx \= xxxx \= \kill
    \query($q:$ \textbf{query}); \\
    1. \> $t$ $\leftarrow$ \textsc{Modify-taxonomy}($q$) \\
    2. \> \emph{ID} $\leftarrow$ \textsc{New-query-id} \\
    3. \> \emph{self}:~\ask(\emph{self}, \emph{ID}, $t,$ $\{t\}$) \\
    4. \> \textbf{wait until} \emph{ID} is closed \textbf{then} \\
    5. \> (\pid, \qid, $t,$ $n,$ $R$) $\leftarrow$ \textsc{Delete}(\emph{ID}) \\
    6. \> \textsc{Cleanup-taxonomy}($t$) \\
    7. \> \textbf{return}($R$)
  \end{tabbing}
}
  \caption{The \query~procedure}
  \label{fig:query}
\end{figure}

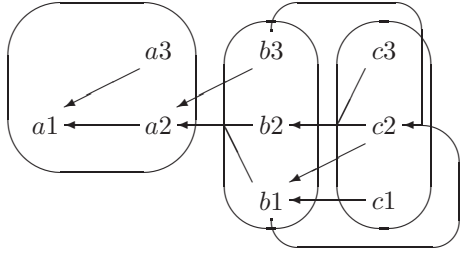
\begin{figure}[htbp]
  \centering
  \begin{picture}(5.75,3.5)
    \put(0,1.5){\makebox(0.5,0.5){$a1$}}
    \put(1.5,1.5){\makebox(0.5,0.5){$a2$}}
    \put(3,1.5){\makebox(0.5,0.5){$b2$}}
    \put(4.5,1.5){\makebox(0.5,0.5){$c2$}}
    \put(1.5,2.5){\makebox(0.5,0.5){$a3$}}
    \put(3,2.5){\makebox(0.5,0.5){$b3$}}
    \put(4.5,2.5){\makebox(0.5,0.5){$c3$}}
    \put(3,0.5){\makebox(0.5,0.5){$b1$}}
    \put(4.5,0.5){\makebox(0.5,0.5){$c1$}}
    \multiput(1.5,2.5)(1.5,0){2}{\vector(-2,-1){1}}
    \multiput(1.5,1.75)(1.5,0){3}{\vector(-1,0){1}}
    \put(3,1){\line(-1,2){0.37}}
    \put(4.5,1.5){\vector(-2,-1){1}}
    \put(4.5,2.5){\line(-1,-2){0.38}}
    \put(4.5,0.75){\vector(-1,0){1}}
    \put(4.25,3){\oval(2,.75)[t]}
    \put(5.25,3){\line(0,-1){1.25}}
    \put(5.25,1.75){\vector(-1,0){.25}}
    \put(4.5,0.5){\oval(2.5,.75)[b]}
    \put(5.25,1.25){\oval(1,1)[tr]}
    \put(5.75,0.5){\line(0,1){.75}}
    \put(1,2.25){\oval(2.5,2.25){}}
    \put(3.25,1.75){\oval(1.25,2.75){}}
    \put(4.75,1.75){\oval(1.25,2.75){}}
  \end{picture}
  \caption{A network taxonomy B-graph}
  \label{fig:netbgraph}
\end{figure}

As an example, let us consider the network shown in Figure~\ref{fig:net}, whose
corresponding B-graph is shown in Figure~\ref{fig:netbgraph}, and the query
$(a2\wedge a3)$ on peer $P_a.$ When given as input to \query, this query is
passed on to \textsc{Modify-taxonomy,} which adds the hyperedge $(\{a2,a3\},t)$
to the taxonomy B-graph and returns the newly generated term $t.$ Let us assume
that $q1$ is the id of the new query. \query~then sends the message \ask($P_a,$
$q1,$ $t,$ $\{t\}$) to itself, and gets into the wait loop until the query is
evaluated.

\subsubsection{\ask}
\label{sec:ask}

For readability, we will describe \ask~and \tell~ as if they were methods whose
parameters are the message fields. \ask~(Figure~\ref{fig:ask}) uses the
following variables:
\begin{itemize}
\item $n:$ counts how many sub-queries the input query \qid~ generates;
\item \qp: is the initial query program of \qid;
\item $Q:$ is a queue holding the information to send the \ask~messages required
  to evaluate \qid;
\item $C:$ is the query sub-program being currently computed.
\end{itemize}
After initialization, \ask~performs (line 2) the same test as \qe, looking for a
hyperedge $h$ in the local B-graph whose head is the given term $t$ and whose tail is disjoint form
$A.$ If no such hyperedge is found, then $n$ remains 0, the test on line 10
fails, and the result of the evaluation of the given term query $t$ is just
$I(t)$ (as \qe~establishes), which \ask~returns by sending a \tell~message to
the invoking peer \pid~(line 15). If instead a hyperedge $h$ is found, then the
intersection of the evaluation of each term $u_i$ in its tail should be added to
the result, according to \qe.  In order to achieve the same behavior,
\ask~enters a loop in which it processes each term $u_i$ to the end of
constructing in $C$ the query sub-program associated to $h.$ First, a new query
id \emph{ID} is generated (line 5) to denote the sub-query on $u_i;$ the newly
generated id is then added to $C.$ On line 7, the number of open calls is
increased by one, and on line 8 the required information to evaluate the query
$u_i$ is enqueued in $Q.$ This information is:
\begin{itemize}
\item the id of the peer $P_h$ holding the terms in the tail of the hyperedge
  $h;$ we assume this information is stored with the hyperedge just for
  convenience, the peer can also store it separately;
\item the \emph{ID} of the sub-query;
\item the query term $u_i$ and
\item the set of the visited terms $A\cup\{u_i\},$ as in \qe.
\end{itemize}
Each sub-program so generated is added to \qp, after considering all relevant
hyperedges (line 9). At this point, if the number of open calls is positive,
\ask~uses the function \textsc{Persist} in order to create the log object
representing the query \qid, and to persist it in the log.  Once the log object
is successfully persisted, \ask~must launch the evaluation of the generated
sub-queries, which it does in the loop on lines 12-14. Until $Q$ is empty, it
dequeues the information for constructing an \ask~message for each sub-query,
and sends such message to the peer $P_h.$ The value of the first message field is
the peer identity (\emph{self}), as the invoking peer.

\begin{figure}
{\small
  \begin{tabbing}
    1\= Les \= xxxx \= xxxx \= xxxx \= xxxx \= \kill
    \ask(\pid,\qid : \textbf{ID}; $t:$ \textbf{term}; $A:$ \textbf{set of terms}); \\
    \> 1. \> $n$ $\leftarrow$ $0;$ ~\qp,~$Q$ $\leftarrow$ $\emptyset$ \\
    \> 2. \> \textbf{for each} hyperedge $h=\langle \{u_1,...,u_r\},t\rangle$ \textbf{such that}
    $\{u_1,...,u_r\}\cap A=\emptyset$ \textbf{do} \\
    \> 3. \> \> $C$ $\leftarrow$ $\emptyset$ \\
    \> 4. \> \> \textbf{for each} $u_i$ \textbf{do} \\
    \> 5. \> \> \> \emph{ID} $\leftarrow$ \textsc{New-query-id} \\
    \> 6. \> \> \> $C$ $\leftarrow$ $C$ $\cup$ \{\emph{ID}\} \\
    \> 7. \> \> \> $n$ $\leftarrow$ $n+1$ \\
    \> 8. \> \> \> \textsc{Enqueue}($Q,$ ($P_h,$ \emph{ID,} $u_i,$ $A\cup\{u_i\}))$ \\
    \> 9. \> \> \qp $\leftarrow$ \qp $\cup$ $\{C\}$ \\
    10. \> \> \textbf{if} $n>0$ \textbf{then} \\
    11. \> \> \> \textsc{Persist}(\pid,\qid,$t,n,$\qp) \\
    12. \> \> \> \textbf{until} $Q\neq\emptyset$ \textbf{do} \\
    13. \> \> \> \> (\emph{P$_h$,ID,u,B}) $\leftarrow$ \textsc{Dequeue}($Q$) \\
    14. \> \> \> \> $P_h:\ask$($\mathit{self,ID},u,B$) \\
    15. \> \> \textbf{else} \pid:\tell(\qid,$I(t))$
  \end{tabbing}
}
  \caption{The procedure to process \ask~messages}
  \label{fig:ask}
\end{figure}

At this point, it can be easily verified that the assumption that all terms in
the tail of a hyperedge are from the same terminology, namely that of peer
$P_h,$ \emph{can be relaxed without any impact on the query evaluation
  procedure.} In logical terms, this is the assumption that the conjunctive
queries on the left-hand side of subsumption relationships are from the query
language of one peer. We have made this assumption because it fits our vision of
a network. But \ask~can easily work also with hyperedges whose tails have terms
from different terminologies: all that is required is to store the id of the peer
holding each term, rather than the id of the peer holding the whole hyperedge.

Let us resume our running example. Upon processing the message ($P_a,$ $q1,$
$t,$ $\{t\}$), \ask~finds that the hyperedge $h=(\{a2,a3\},t)$ passes the test
on line 2, and enters the loop on the tail of $h.$ For term $a2,$ assuming the
generated query id is $q2,$ the record ($P_a,$ $q2,$ $a2,$ $\{t,a2\}$) is
enqueued in $Q,$ while for term $a3,$ (generated id $q3$) it is enqueued the
record ($P_a,$ $q3,$ $a3,$ $\{t,a3\}$). As there are no more hyperedges and
$n=2,$ a new log object is created to represent the query $t.$ The attributes of
this object are:
\bl
\item \pid~= $P_a$
\item \qid~= $q1$
\item $t~=~t$
\item $n~=~2$
\item \qp~=~$\{\{q2,q3\}\}.$
\el
Now two \ask~messages are send to $P_a:$
\begin{enumerate}
\item ($P_a,$ $q2,$ $a2,$ $\{t,a2\}$), and
\item ($P_a,$ $q3,$ $a3,$ $\{t,a3\}$).
\end{enumerate}
Let us see how the latter message is processed. Since there are no incoming
hyperedges into term $a3,$ $n$ remains 0, and the processing of the message is
concluded by the sending of the message \tell($q3,$ $I(a3)$) to $P_a.$

\subsubsection{\tell}
\label{sec:tell}

When a peer receives a \tell(\qid,$R$) message (see Figure~\ref{fig:tell}),
\qid~is an open call of some log object in the peer's log, in the program of
some term (sub)query $t$ with id \qidone. Then, as a first action, the peer retrieves this object by
using the \textsc{Delete1} function, which takes as input \qid, returns the
object and \emph{deletes} it form the log.  Notice that there is exactly one
object having \qid~as open call, since \ask~generates a new id for each
sub-query it identifies, as we have already seen. After retrieving the log
object, \tell~uses \textsc{Close} to modify the query program \qp~in it, by
closing the open call \qid: this means to replace \qid~by $R,$ obtaining a new
query program \qpone. On line 3, the number of open calls of the log object is
tested: if it is 1, then the just closed call was the last one to be open in
query \qidone; in this case, the result of \qidone~ is computed in $S$ by
\textsc{Compute-answer}. For a given program:
\[
\mathit{QP} = \{\subp_1,\ldots,\subp_m\}
\]
where each sub-program $\subp_j$ is given by a collection of object sets:
\[
\subp_j = \{R^j_1,\ldots,R^j_{m_j}\}
\]
\textsc{Compute-answer} returns:
\[
S=\bigcup \{ \bigcap \subp_j~|~1\leq j\leq m\}
\]
$S \cup I(t)$ is exactly what the \qe~procedure computes.
If $t$ is not in the terminology of the peer $(t \not \in T_{self})$ then it follows that
\qidone~is the id of the original query $q$. Thus, $I(t)=\emptyset$ and $S=\ans(t,S_\N)$.
Therefore, the object (\pid, \qidone, $t, 0, S$)
 is persisted in the log (line 5), indicating to
\textsc{Query}($q$) (Figure \ref{fig:query}) that the evaluation of the query $q$ has finished.
Otherwise, the so obtained result $S \cup I(t)$
is \tell-ed to
the peer \pid~which, according to the log object, was the one to \ask~the
evaluation of \qidone. Notice that this may fire another \tell~message, in case
\qidone~is the last open call of some other query. If the test on line 3 fails,
then there are still open calls in the log object, which is therefore persisted
back by \textsc{Persist} on line 6, after decreasing the number of open calls in
it and replacing the query program \qp~by the updated one \qpone.

\begin{figure}[htbp]
{\small
  \begin{tabbing}
    Les \= xxxx \= xxxx \= xxxx \= xxxx \= \kill
    \tell(\qid: \textbf{ID}; $R:$ \textbf{set of objects}); \\
    1. \> (\pid, \qidone, $t,$ $n,$ \qp) $\leftarrow$ \textsc{Delete1}(\qid) \\
    2. \> \qpone $\leftarrow$ \textsc{Close}(\qp, \qid, $R$) \\
    3. \> \textbf{if} $n=1$ \textbf{then} \\
    4. \> \> $S$ $\leftarrow$ \textsc{Compute-answer}(\qpone) \\
    5. \> \>  \textbf{if} $t \not \in T_{self}$ \textbf{then} \textsc{Persist}(\pid, \qidone, $t, 0, S$)\\
    6. \> \>   \textbf{else} \pid:\tellc(\qidone, $t, S\cup I(t)$) \\
    7. \> \textbf{else} \textsc{Persist}(\pid, \qidone, $t,$ $n-1,$ \qpone)
  \end{tabbing}
  }
  \caption{The procedure to process \tell~messages}
  \label{fig:tell}
\end{figure}

In our example, the message  \tell($q3,$ $I(a3)$) is received by peer $P_a.$ The
function \textsc{Delete1} returns the log object ($P_a,$ $q1,$ $t,$ $2,$
$\{\{q2,q3\}\}$), the only one that has the open call $q3.$ \textsc{Close}
produces the new query program $\{\{q2,I(a3)\}\},$ and since $n$ is not 1, the
following modified log object is persisted:
\begin{center}
($P_a,$ $q1,$ $t,$1,$\{\{q2,I(a3)\}\}$).
\end{center}

The example is completed in appendix.

\subsection{Correctness and complexity}
\label{sec:cc}

As it has been argued, the combined action of the procedures processing \ask~and
\tell~messages is equivalent to the behavior of the procedure \qe. To see
why in more detail, it suffices to consider the following facts:
\begin{enumerate}
\item An \ask~message is generated for each recursive call performed by \qe~and
  vice-versa, that is whenever \qe~would perform a recursive call, an
  \ask~message is generated.  This is guaranteed by the fact that the test on
  line 2 of \ask~is the same as the test on line 3 of \qe. Therefore, the
  number of \ask~messages is the same as the number of terms that can be found
  on a B-path from $t.$
\item For each \ask~message, at most one log object is generated and persisted.
\item For each \ask~message, a \tell~message results, and no more. This can be
  observed by considering that, for each processed \ask~message, there can be
  two cases:
  \begin{enumerate}
  \item no hyperedge is found that passes the test on line 2 of \ask: in this
    case, no subsequent \ask~message is generated, and a \tell~message is
    generated;
  \item at least one hyperedge passes the test: in this case a number of
    sub-queries is generated and registered in the query program of the log
    object. Each such sub-query is evaluated by issuing an \ask~message with a
    larger set of visited terms. Since the B-graph is finite, eventually each
    sub-query will lead to a term falling in the previous case (this is how
    \qe~terminates). When all sub-queries of a given term query $t$ are closed, the
    number of open calls of $t$ goes down to 0, and \tell~issues another
    \tell~message on $t.$ This will propagate closure up, until all open calls
    are closed.
  \end{enumerate}
\item Finally, the \textsc{Compute-answer} procedure performs the same operation
  on the result of sub-queries as \qe~does on the results of its recursive
  calls.
\end{enumerate}

As a consequence of these facts we have the correctness of the network query
evaluation procedure, and also its efficiency. In fact, the total number of
messages generated is twice the number of terms visited by \qe, and the number
of log objects is no larger than that.

\section{Optimization issues}
\label{sec:opt}

So far, we have focused on correctness.  In this Section we discuss
optimization.  There are many techniques that are potentially useful to this
end. For instance, when sub-queries return large results, their closing
(performed by \textsc{Close}) and the computation of their results
(\textsc{Compute-answer}) should be done with care.  However, dealing with all
the relevant optimization techniques goes beyond the scope of this paper.
Instead, we focus on caching (Section \ref{sec:cache}), which is applicable to
all situations, and on exploiting data structures employed in structured P2P
systems, namely Distributed Hash Tables (like in Chord \cite{Stoica01}). This
latter issue is tackled in Sections \ref{sec:DHT_O} and \ref{sec:DHT_T};
besides
showing how to further improve the efficiency of the system, the ensuing
discussion hints at how to extend the applicability of our model, and highlights
the relationship with a large part of the literature on P2P systems. More on
related work can be found in Section~\ref{sec:rw}.

\subsection{Caching}
\label{sec:cache}

A strong point of our model is that the adoption of caches could significantly
speed up the evaluation of queries, by reducing both the latency time and the
network throughput. This is because the set of queries that a peer can send to
its articulated peers is bounded in size and can be pre-determined: it comprises
all ``foreign'' queries of the peer, \ie~queries that appear as left-hand sides
in the peer's articulations. Note that the number of queries that a peer can
propagate to its neighbors is unbounded in other models of P2P systems, for
example in Gnutella, where each peer propagates whatever query it
receives\footnote{FreeNet tries to improve the situation by forwarding queries
  (and new objects too) only to those peers that, according to the contents of
  the cache, have similar keys.  In this way, each cache tends to have entries
  about similar keys and this tends to improve the quality of routing over time.
}.
It follows that the caches of our model will enjoy higher hit ratios
compared to other P2P models, for the same cache size.
The subsequent subsections present three caching policies, namely:
\begin{itemize}
    \item caching answers of local terms,
    \item caching answers of local terms and pushing answers of articulation tails, and
    \item caching answers of articulation heads.
\end{itemize}

\subsubsection{Caching  answers of local terms}

According to this caching policy, each peer $\ES$ caches pairs of the
form $(t,\ans(t,S_\N))$, where $t$ is a term in the peer's terminology $T_\ES.$
If there are no memory limitations for caches, then after a while each peer will
have cached its whole terminology, and query evaluation reduces to locally
calculating the extension of the query by union-ing and intersecting the
extensions of the peer's terms. In other words, any peer will be able to
evaluate network queries over its own taxonomy without sending any message to
the network\footnote{Apart those required for re-evaluating queries when updates
  occur.}!
This is of course the idealistic case. In general, only some terms (possibly
none) will be cached in each peer. Under these circumstances, when a peer $\ES$ receives
an $\ask$ message for a term query $t$, the \ask~procedure checks which of the answers for
the term (sub)queries needed for the evaluation of $t$
are in the cache, and issues \ask~messages only for evaluating the remaining
terms.

The modified query evaluation algorithms for supporting this caching policy
are parts of the algorithms for the more general policy
 that is described in Section \ref{sec:CachePush}.

\subsubsection{Caching answers of local terms and pushing answers of articulation tails}
\label{sec:CachePush}

A complementary scenario,
best suited  for a P2P system that offers
recommendation services in push-style manner,
is to assume that each peer $\ES$ knows also the articulations
$t_1\wedge\ldots\wedge t_r\preceq u$ from
other peers $\ES'$ to $\ES$ (called {\em foreign articulations}).
In this case, if all the terms $t_1,\ldots, t_r$ are cached
in $\ES,$ then $\ES$ can send to $\ES'$ the pair $(t_1\wedge\ldots\wedge
t_r,\ans(t_1\wedge\ldots\wedge t_r, S_\N))$ to be stored in the cache of $\ES'.$
This can be done because  from Proposition~\ref{prop:MinModel}
and Definition \ref{def:QExt} it follows that
\[
\ans(t_1\wedge\ldots\wedge t_r, S_\N) = \bigcap \{\ans(t_i,S_\N)~|~1\leq i\leq
r\}
\]

The cache is exploited by the modified
\ask~procedure (\askc), shown in Figure~\ref{fig:askc}.  The
modified with caching \tell~procedure (\tellc) is shown in Figure~\ref{fig:tellc}. The modifications are indicated by bold line numbers and are described
in a semi-formal way, in order to abstract from irrelevant details.

\begin{figure}
{\small
  \begin{tabbing}
    1\= Les \= xxxx \= xxxx \= xxxx \= xxxx \= xxxx \= xxxx\kill
    \askc(\pid,\qid : \textbf{ID}; $t:$ \textbf{term}; $A:$ \textbf{set of terms}); \\
    \> {\bf 1.} \> \textbf{if} $t$ is cached \textbf{then}  \pid:\tellc($\qid, t, \ans(t,S_\N)$)\\
    \> {\bf 2.} \> \textbf{else} \textbf{if} $|A|=2$ \textbf{then} add $t$ into  \textsc{to-be-cached} log  //
                    {\small $t$ is a term of the original query $q$}\\
    \> 3. \> \> $n$ $\leftarrow$ $0;$ ~\qp, $Q,$ $S$ $\leftarrow$ $\emptyset$ \\
    \> 4. \> \> \textbf{for each} hyperedge $h=\langle \{u_1,...,u_r\},t\rangle$ \textbf{such that}
    $\{u_1,...,u_r\}\cap A=\emptyset$ \textbf{do} \\
    \> {\bf 5.} \> \> \> \textbf{if} $P_h \not = self$ and $u_1\wedge\ldots\wedge u_r$ is cached \textbf{then} $C$ $\leftarrow$
    \{$\ans(u_1\wedge\ldots\wedge u_r,S_\N)$\} \\
    \> 6. \> \> \> \textbf{else} $C$ $\leftarrow$ $\emptyset$ \\
    \> {\bf 7.} \> \> \> \> \textbf{for each} $u_i$ \textbf{do} \\
    \> {\bf 8.} \> \> \> \> \>   \textbf{if} $P_h  = self$  and $u_i$ is cached \textbf{then}\\
    \> {\bf 9.} \> \> \> \> \> \>  $C$ $\leftarrow$ $C$ $\cup$ $\{\ans(u_i, S_\N)\}$ \\
    {\bf 10.} \> \> \> \> \> \>   \textbf{else}\\
    11. \> \> \> \> \> \> \> \emph{ID} $\leftarrow$ \textsc{New-query-id} \\
    12. \> \> \> \> \> \> \> $C$ $\leftarrow$ $C$ $\cup$ \{\emph{ID}\} \\
    13. \> \> \> \> \> \> \> $n$ $\leftarrow$ $n+1$ \\
    14. \> \> \> \> \> \> \> \textsc{Enqueue}($Q,$ ($P_h,$ \emph{ID,} $u_i,$ $A\cup\{u_i\}))$ \\
    15. \> \> \> \> \qp $\leftarrow$ \qp $\cup$ $\{C\}$ \\
    16. \> \> \> \textbf{if} $n>0$ \textbf{then} \\
    17. \> \> \> \> \textsc{Persist}(\pid,\qid,$t,n,$\qp) \\
    18. \> \> \> \> \textbf{until} $Q\neq\emptyset$ \\
    19. \> \> \> \> \> (\emph{P$_h$,ID,u,B}) $\leftarrow$ \textsc{Dequeue}($Q$) \\
    20. \> \> \> \> \> $P_h$:\askc($\mathit{self,ID},u,B$) \\
    {\bf 21.} \> \> \>  \textbf{else if} \qp$\neq\emptyset$ \textbf{then} $S$ $\leftarrow$
    \textsc{Compute-answer}(\qp)  \\
    22. \> \> \> \> \pid:\tellc($\qid, t, S\cup I(t)$)
  \end{tabbing}
}
  \caption{The procedure to process \ask~messages with cache}
  \label{fig:askc}
\end{figure}

\begin{figure}[htbp]
{\small
  \begin{tabbing}
    1\= Les  \= xxxx \= xxxx \= xxxx \= xxxx \= xxxx \=xxxx \kill
    \tellc(\qid: \textbf{ID}; $t':$ \textbf{term}; $R:$ \textbf{set of objects}); \\
    \> {\bf 1.} \> \textbf{if} $t'$ in \textsc{to-be-cached} log \textbf{then}
                                //  {\small $t'$ is a term of the original query $q$}\\
    \> {\bf 2.} \> \> delete $t'$ from \textsc{to-be-cached} log \\
    \> {\bf 3.} \> \> \textsc{CACHE}$(t', R)$\\
    \> {\bf 4.} \> \> \textbf{for each} foreign articulation  $t_1\wedge\ldots\wedge t_r \preceq u$ from another peer $\ES$
    to $self$  \textbf{do}\\
    \> {\bf 5.} \> \> \> \textbf{if} $t' \in \{t_1, ..., t_r\}$  and all $t_1$, ..., $t_r$ are cached \textbf{then}\\
    \> {\bf 6.} \>\> \> \> forward to $\ES$ the pair $(t_1\wedge\ldots\wedge t_r, \ans(t_1\wedge\ldots\wedge t_r,S_\N)$ for
    caching\\
    \> 7. \> (\pid, \qidone, $t,$ $n,$ \qp) $\leftarrow$ \textsc{Delete1}(\qid) \\
    \> 8. \> \qpone $\leftarrow$ \textsc{Close}(\qp, \qid, $R$) \\
    \> 9. \> \textbf{if} $n=1$ \textbf{then} \\
    10. \> \> \> $S$ $\leftarrow$ \textsc{Compute-answer}(\qpone) \\
    11. \> \> \> \textbf{if} $t \not \in T_{self}$ \textbf{then} \textsc{Persist}(\pid, \qidone, $t, 0, S$)\\
    12. \> \> \>  \textbf{else} \pid:\tellc(\qidone, $t, S\cup I(t)$) \\
    13. \> \> \textbf{else} \textsc{Persist}(\pid, \qidone, $t,$ $n-1,$ \qpone)
  \end{tabbing}
}
  \caption{The procedure to process \tell~messages with cache}
  \label{fig:tellc}
\end{figure}

The cache of a peer $\ES$ consists of two kinds of pairs:
\begin{itemize}
\item $(t',\ans(t',S_\N))$ where $t'$ is a term in the peer's terminology $T_\ES.$
  Pairs of this kind are inserted into the cache by the \tellc($QID,t',R)$~procedure\footnote{Note that
  \tellc($QID,t',R)$ takes an extra argument $t'$, which is the term query corresponding to query id $QID$.}, when the
  peer $\ES$ is
  \tell-ed the answer $R$ for a term query $t'$, initiated by an \ask~message of type
  \begin{center}
    \ask($\pid,$ $\qid,$ $t',$ $\{u,t'\}$)
  \end{center}
  where $u$ is a new term created by \query($q)$~to represent the original (complex) query $q$, posed to peer
  $\ES$. This means
  that the term $t'$ appears in $q$ and is not evaluated in
  the context of the evaluation of a more general term. For example, this is the case of the
  \ask~messages presented at the end of Section~\ref{sec:ask}:
  \begin{enumerate}
  \item ($P_a,$ $q2,$ $a2,$ $\{t,a2\}$), and
  \item ($P_a,$ $q3,$ $a3,$ $\{t,a3\}$).
  \end{enumerate}
  In this way, based on the correctness of the \textsc{query} procedure (Section \ref{sec:cc}),
  it is guaranteed that $R=\ans(t',S_\N)$, \ie~the received answer $R$ is the full answer for $t'$
  and not a subset of
  it, reduced due to cycles in the taxonomy $(T_\N, \preceq_\N)$. Thus, the pair $(t',R)$ can be safely cached.

\item $(t_1\wedge\ldots\wedge t_r,\ans(t_1\wedge\ldots\wedge t_r, S_\N))$ where
  $t_1\wedge\ldots\wedge t_r\preceq u$ is an articulation from $\ES$ to $\ES',$
  \ie~$u\in T_\ES$ and $t_1,\ldots,t_r\in T_{\ES'}.$ Each such pair
  is forwarded to $\ES$ by the \tellc~procedure executed at the peer $\ES'$,
  upon realizing that all the terms involved in
  the left-hand side of the articulation are stored in the local (to $\ES'$) cache.
  In particular, this check is made immediately after a pair $(t', \ans(t',S_\N))$ is
  added in the cache of $\ES'$, where $t' \in \{t_1,...,t_r\}$
  (see lines 3-6 of \tell$_c$).
\end{itemize}

Below are
the main differences of \askc$(\pid,\qid,t,A)$
with respect to the cache-less \ask:
\begin{itemize}
  
\item If the answer to the term query $t$ \ask-ed by peer $PID$ is in the cache,
  then the answer is immediately \tell-ed to peer $PID$. Otherwise, if $|A|=2$
  then $t$ is added in the \textsc{to-be-cached} log ($t$ is a term of the
  original query $q$).  The \textsc{to-be-cached} log is checked by
  \tellc$(QID,t',R)$.  If $t'$ is found in the \textsc{to-be-cached} log then
  $(t', R)$ is added to the local cache through the \textsc{cache}$(t', R)$
  command (line 3 of \tellc).
  
\item Before processing the tail of a hyperedge $h$ which passes the test on
  line 4, a test is performed, to ascertain whether the query corresponding to
  the tail, given by $u_1\wedge\ldots\wedge u_r,$ is in the cache (this test is
  needed only if $P_h \not = self$, \ie~$h$ corresponds to an articulation
  hyperedge).  If yes, the only action taken is the insertion of
  $\ans(u_1\wedge\ldots\wedge u_r,S_\N)$ into the query sub-program \qp~being
  built (line 15). If the query is not in the cache, then for each $u_i$, it is
  checked if its answer is in the cache (this test is needed only if $P_h =
  self$). If not, then the execution proceeds normally.
  
\item If all sub-queries are cached, then when all relevant hyperedges have been
  processed (line 16), $n$ is zero but \qp~is not empty. In this case the test
  on line 21 is passed, and the result of \qid~is computed in $S$ as if closing
  \qp~in a \tell. $S$ is subsequently returned along with $I(t).$ If \qp~is
  empty, then no hyperedge has been found and $S=\emptyset$. So, the result
  returned to the user is simply $I(t).$
\end{itemize}

We would like to note that our algorithms can further be extended such that
\tellc~caches the answer $S \cup I(t)$ for term sub-queries $t$ before \tell-ing
them to the requesting peer $PID$ (line 12 of \tellc), as long as it is certain
that $S \cup I(t)=\ans(t,S_\N)$.  This is the case if (i) for each term $u$ of a
peer $\ES'$ encountered during the evaluation of $t$ (including $t$ itself), all
hyperedges $\langle \{u_1,...,u_r\},u\rangle$ of the taxonomy B-graph of $\ES'$
pass the test of line 4 of \askc, or (ii) $u$ is cached.  Thus, (i) no
evaluation path of $u$ is eliminated due to cycles in the taxonomy $(T_\N,
\preceq_\N)$ or (ii) $\ans(u,S_\N$) is immediately retrieved from the cache.

For this reason \textsc{Persist}(\pid,\qid,$t,n,$\qp) and \tellc(\qid,$t',R)$
should be extended with an extra field $flag$ that takes the values {\tt full}
or {\tt partial}.  A (query) log object (\pid,\qid,$t,n,\qp,flag)$, where
$flag=${\tt full}, of a peer $\ES$ indicates that (i) for all {\em closed} term
sub-queries of $QP$, full answers have been received and (ii) all hyperedges
$\langle \{u_1,...,u_r\},t\rangle$ of the taxonomy B-graph of $\ES$ have passed
the test of line 4 of \askc. If this is not the case, $flag=${\tt partial}.  A
message \tellc$(QID,t',R, flag)$, where $flag=${\tt full}, indicates that
$R=\ans(t',S_\N)$, whereas a message \tellc$(QID,t',R, flag)$, where $flag=${\tt
  partial}, indicates that $R \subseteq \ans(t',S_\N)$.  Thus, based on the
$flag$ information, the \tellc~procedure executed at a peer will always be able
to know if the computed answer $S \cup I(t)$ for a term sub-query $t$ requested
by peer $PID$ is a full or partial answer. In the case of a full answer and if
$t$ is the head of an articulation hyperedge then $(t, S \cup I(t))$ is cached.
We want to note that the latter condition is not a strong condition and is
needed only in order to reduce the cache size, while taking the most advantage
of caching.

The extended \askc~procedure (\askce) and the extended \tellc~procedure
(\tellce) are given in Figures \ref{fig:askce} and \ref{fig:tellce},
respectively. The modifications are indicated by bold line numbers.  Note that
\tellce~calls the procedure \textsc{Cache\&Forward} (Figure
\ref{fig:CacheForward}), when a pair $(t, \ans(t,S_\N))$ is going to be stored
in the cache. Additionally, \tellce~uses the function $min(flag,flag')$ (lines
8, 11), which returns the minimum of the flag values $flag$, $flag'$, based on
the ordering {\tt partial} $\leq$ {\tt full}. This guarantees that the flag
value of the \tellce~message in line 8 and the log object in line 11 is correct.

\begin{figure}
{\small
  \begin{tabbing}
    1\= Les \= xxxx \= xxxx \= xxxx \= xxxx \= xxxx \= xxxx \= xxxx \=xxxx\kill
    \askce(\pid,\qid : \textbf{ID}; $t:$ \textbf{term}; $A:$ \textbf{set of terms}); \\
    \> {\bf 1.} \> \textbf{if} $t$ is cached \textbf{then}  \pid:\tellce($\qid, t, \ans(t,S_\N)$, {\tt full})\\
    \> 2. \> \textbf{else} \textbf{if} $|A|=2$ \textbf{then} add $t$ into  \textsc{to-be-cached} log  //
                    {\small $t$ is a term of the original query $q$}\\
    \> {\bf 3.} \> \> $n$ $\leftarrow$ $0;$ ~\qp, $Q,$ $S$ $\leftarrow$ $\emptyset$; $flag=${\tt full} \\
    \> 4. \> \> \textbf{for each} hyperedge $h=\langle \{u_1,...,u_r\},t\rangle$  \textbf{do} \\
    \> 5. \> \> \> \textbf{if} $\{u_1,...,u_r\}\cap A=\emptyset$ \textbf{then} \\
    \> 6. \> \> \> \> \textbf{if} $P_h \not = self$ and $u_1\wedge\ldots\wedge u_r$ is cached \textbf{then} $C$ $\leftarrow$
    \{$\ans(u_1\wedge\ldots\wedge u_r,S_\N)$\} \\
    \> 7. \> \> \> \> \textbf{else} $C$ $\leftarrow$ $\emptyset$ \\
    \> 8. \> \> \> \> \> \textbf{for each} $u_i$ \textbf{do} \\
    \> 9. \> \> \> \> \> \>  \textbf{if} $P_h  = self$  and $u_i$ is cached \textbf{then}\\
    10. \>  \> \> \> \> \> \>  \> $C$ $\leftarrow$ $C$ $\cup$ $\{\ans(u_i, S_\N)\}$ \\
    11. \> \> \> \> \> \>  \> \textbf{else}\\
    12. \> \> \> \> \> \> \> \> \emph{ID} $\leftarrow$ \textsc{New-query-id} \\
    13. \> \> \> \> \> \> \> \> $C$ $\leftarrow$ $C$ $\cup$ \{\emph{ID}\} \\
    14. \> \> \> \> \> \> \> \> $n$ $\leftarrow$ $n+1$ \\
    15. \> \> \> \> \> \> \> \> \textsc{Enqueue}($Q,$ ($P_h,$ \emph{ID,} $u_i,$ $A\cup\{u_i\}))$ \\
    16. \> \> \> \> \> \qp $\leftarrow$ \qp $\cup$ $\{C\}$ \\
    {\bf 17.} \> \> \> \> \textbf{else} $flag=${\tt partial}\\
    18. \> \> \>  \textbf{if} $n>0$ \textbf{then} \\
    {\bf 19.} \> \> \> \> \textsc{Persist}(\pid, \qid, $t$, $n$, \qp, $flag$) \\
    20. \> \> \> \> \textbf{until} $Q\neq\emptyset$ \\
    21. \> \> \> \> \> (\emph{P$_h$,ID,u,B}) $\leftarrow$ \textsc{Dequeue}($Q$) \\
    22. \> \> \> \> \> $P_h$:\askce($\mathit{self,ID},u,B$) \\
    23. \> \> \>  \textbf{else if} \qp$\neq\emptyset$ \textbf{then} $S$ $\leftarrow$
    \textsc{Compute-answer}(\qp)  \\
    {\bf 24.} \> \> \> \> \pid:\tellce(\qid, $t$, $S\cup I(t)$, $flag$)
  \end{tabbing}
  }
  \caption{The extended procedure to process \ask~messages with cache}
  \label{fig:askce}
\end{figure}

\begin{figure}[htbp]
{\small
  \begin{tabbing}
    1\= Les  \= xxxx \= xxxx \= xxxx \= xxxx \= xxxx \=xxxx \kill
    \tellce(\qid: \textbf{ID}; $t':$ \textbf{term}; $R:$ \textbf{set of objects}; $flag': \{${\tt full}, {\tt partial}$\}$); \\
    \> 1. \> \textbf{if} $t'$ in \textsc{to-be-cached} log \textbf{then}
                                //  {\small $t'$ is a term of the original query $q$}\\
    \> 2. \> \> \textsc{Cache\&Forward}$(t', R)$\\
    \> {\bf 3.} \> (\pid, \qidone, $t,$ $n,$ \qp, $flag$) $\leftarrow$ \textsc{Delete1}(\qid) \\
    \> 4. \> \qpone $\leftarrow$ \textsc{Close}(\qp, \qid, $R$) \\
    \> 5. \> \textbf{if} $n=1$ \textbf{then} \\
    \> 6. \> \> $S$ $\leftarrow$ \textsc{Compute-answer}(\qpone) \\
    \> {\bf 7.} \> \> \textbf{if} $t \not \in T_{self}$ \textbf{then} \textsc{Persist}(\pid, \qidone, $t$, 0, $S$, {\tt full})\\
    \> {\bf 8.} \> \>  \textbf{else} \pid:\tellce(\qidone, $t$, $S\cup I(t)$, $min(flag,flag')$) \\
    \> {\bf 9.} \> \> \>  \textbf{if} $min(flag,flag')$={\tt full} and $t$ is the head of an articulation hyperedge \textbf{then} \\
    {\bf 10.} \> \> \> \> \> \textsc{Cache\&Forward}$(t, S\cup I(t))$\\
    {\bf 11.} \> \> \textbf{else} \textsc{Persist}(\pid, \qidone, $t,$ $n-1$,  \qpone, $min(flag,flag')$)
  \end{tabbing}
}
  \caption{The extended procedure to process \tell~messages with cache}
  \label{fig:tellce}
\end{figure}

\begin{figure}[htbp]
 {\small
  \begin{tabbing}
    1\= Les  \= xxxx \= xxxx \= xxxx \= xxxx \= xxxx \=xxxx \kill
    \textsc{Cache\&Forward}($t:$ \textbf{term}; $R:$ \textbf{set of objects}); \\

    \>   \small{ // It stores the pair $(t,R)$ in the local cache and checks if related (foreign articulation)}\\
    \>\> \small{query-answer pairs can be forwarded to other peers  for caching}\\
    \> 1. \> \> \textsc{CACHE}$(t, R)$\\
    \> 2. \> \> \textbf{if} $t$ in \textsc{to-be-cached} log \textbf{then} delete $t$ from \textsc{to-be-cached}\\
    \> 3. \> \> \textbf{for each} foreign articulation  $t_1\wedge\ldots\wedge t_r \preceq u$ from another peer $\ES$
    to $self$  \textbf{do}\\
    \> 4. \> \> \> \textbf{if} $t \in \{t_1, ..., t_r\}$ and all $t_1$, ..., $t_r$ are cached \textbf{then}\\
    \> 5. \>\> \> \> forward to $\ES$ the pair $(t_1\wedge\ldots\wedge t_r, \ans(t_1\wedge\ldots\wedge t_r,S_\N)$ for
    caching

  \end{tabbing}
 }
  \caption{The procedure \textsc{Cache\&Forward}}
  \label{fig:CacheForward}
\end{figure}

\begin{figure}
{\small
  \begin{tabbing}
    1\= Les \= xxxx \= xxxx \= xxxx \= xxxx \= xxxx \= xxxx\kill
    \askca(\pid,\qid : \textbf{ID}; $t:$ \textbf{term}; $A:$ \textbf{set of terms}); \\
    \> 1. \> \textbf{if} $t$ is cached \textbf{then}  \pid:\tell($\qid, t, \ans(t,S_\N)$)\\
    \> 2. \> \textbf{else} $n$ $\leftarrow$ $0;$ ~\qp, $Q,$ $S$ $\leftarrow$ $\emptyset$ \\
    \> 3. \> \> \textbf{for each} hyperedge $h=\langle \{u_1,...,u_r\},t\rangle$ \textbf{such that}
    $\{u_1,...,u_r\}\cap A=\emptyset$ \textbf{do} \\
    \> 4. \> \> \>  $C$ $\leftarrow$ $\emptyset$ \\
    \> 5. \> \> \> \textbf{for each} $u_i$ \textbf{do} \\
    \> 6. \> \> \>  \>   \textbf{if} $u_i$ is cached \textbf{then}\\
    \> 7. \> \> \>  \> \>  $C$ $\leftarrow$ $C$ $\cup$ $\{\ans(u_i, S_\N)\}$ \\
    \> 8. \> \> \> \>  \>   \textbf{else}\\
    \> 9. \> \> \> \> \> \>  \emph{ID} $\leftarrow$ \textsc{New-query-id} \\
    10. \> \> \> \> \> \>  $C$ $\leftarrow$ $C$ $\cup$ \{\emph{ID}\} \\
    11. \> \> \> \> \> \>  $n$ $\leftarrow$ $n+1$ \\
    12. \> \> \> \> \> \>  \textsc{Enqueue}($Q,$ ($P_h,$ \emph{ID,} $u_i,$ $A\cup\{u_i\}))$ \\
    13. \> \> \>  \> \qp $\leftarrow$ \qp $\cup$ $\{C\}$ \\
    14. \> \> \> \textbf{if} $n>0$ \textbf{then} \\
    15. \> \> \> \> \textsc{Persist}(\pid,\qid,$t,n,$\qp) \\
    16. \> \> \> \> \textbf{until} $Q\neq\emptyset$ \\
    17. \> \> \> \> \> (\emph{P$_h$,ID,u,B}) $\leftarrow$ \textsc{Dequeue}($Q$) \\
    18. \> \> \> \> \> $P_h$:\askca($\mathit{self,ID},u,B$) \\
    19. \> \> \>  \textbf{else if} \qp$\neq\emptyset$ \textbf{then} $S$ $\leftarrow$
    \textsc{Compute-answer}(\qp)  \\
    20. \> \> \> \> \pid:\tell($\qid, S\cup I(t)$)
  \end{tabbing}
}
  \caption{An alternative procedure to process \ask~messages with cache}
  \label{fig:askcalt}
\end{figure}

\subsubsection{Caching answers of articulation heads}

The previous algorithms will cache the most frequently used terms, taking full advantage of caching
with no extra cost for computing cached answers.
However, caches may get filled very quickly.
Below we investigate the case that we cache only the heads
of articulation hyperedges, as the cached answer of these terms is the most beneficial for speeding-up query
answering.
For instance, in the example of Figure  \ref{fig:netbgraph}, we want to cache only $a_2$ on Peer $P_a$,
$b_1$ and $b_2$ on Peer $P_b$, and  $c_2$ on Peer $P_c$.

For this alternative caching case, a top algorithm can be easily designed such
that whenever a peer receives an external query $q$, it finds the local terms
that are heads of articulation hyperedges and are needed for the evaluation of
the query. Then, for each such term $t$, if $t$ is not cached, it calls the
\textsc{Query}($t$) procedure (Figure \ref{fig:query}) and it caches $t$ along
with the received answer $R$, as it is certain that $R=\ans(t,S_\N)$.  This will
fill the needed caches.  The answer of the original query is then computed
locally (\eg by a version of the \qe~procedure, modified with caching).  Note
that \textsc{Query}($t$), in this case, should call \askca~(Figure
\ref{fig:askcalt}) which is a simplified version of \askc~that issues \askca~and
\tell~messages.  Though this approach has the extra cost of requiring full
answers for terms that do not belong to the original query $q$, it is the most
beneficial with respect to the trade-off cache size versus speed.

Of course, another alternative is if the above mentioned top algorithm asks for
the answers of foreign terms $t$ (through \textsc{Query}($t$)) that appear in
the body of articulation hyperedges, instead of asking for the answers of
(local) terms $t$ that are heads of articulation hyperedges.

\subsubsection{Synopsis}

Above we described three caching policies.  Overall, four query evaluation modes
can be supported by our model.  The three caching policies result in faster
query evaluation, but possibly not very updated results, since taxonomies,
interpretations and articulations change.  The mode without cache results in
fresher results but with a slower query evaluation.

In case there are memory limitations for caches, various update policies could
be employed, \eg~keep in cache only the answers of the most frequently used
terms, or keep in cache only some parts of the answers, for instance ``popular''
objects according to some external information collected for this purpose
(object-ranking techniques similar to page-ranking techniques for the Web could
be employed to this end).

\subsection{Querying for object descriptions}
\label{sec:DHT_O}

The query language of our model is term-centered, in the sense that users can
extract information from a source only by asking (Boolean combinations of)
terms. But sometimes it would be useful for the user to better understand the
contents of an object, or the meaning or usage of terms. In these cases, a user
would like to be able to ask ``what are the terms that are used for describing
this object?''  This question can be modulated in different ways, depending
whether or not only local terms are desired, and whether or not only most
specific terms are desired. Correspondingly, an enhanced query language would
offer 4 types of queries, for a given object $o:$
\begin{itemize}
\item the most specific, local terms describing $o;$ assuming the local source
  is $S=(T,\preceq,\obj,I),$ the semantics of this query would be $\indo;$
\item the local terms describing $o,$ that is $\{t\in T~|~o\in\ans(t,S)\};$
\item the most specific terms describing $o$ in the network; assuming $\N$ is
  the network, this query would return
  $\bigcup\{\mathit{ind}_{\ES_i}(o)~|~\ES_i\in\N\};$
\item the terms describing $o$ in the network, that is $\bigcup\{t\in
  T_\N~|~o\in\ans(t,S_\N)\}.$
\end{itemize}
The last two queries clearly make sense only if the objects are shared amongst
the peers, otherwise their results would be the same as that of the previous
two, respectively.

Assuming the peers are willing to share their interpretation, an efficient way
of answering queries of these kinds would be to ``invert the network'', that is
to assign each object $o$ to one peer\footnote{as opposed to assign each term
  $t$ to one peer.}. The designated peer can store all terms that have been
assigned to $o$ by any peer of the network, \ie~ $\mathit{ind}_{S_\N}(o)$.
Interestingly, much work on P2P systems has focused on the design of data
structures for solving this kind of problems (see Section \ref{sec:rw}).  The existence
of a Distributed Hash Table (DHT) as an additional data structure (considering
$\obj$ as the set of keys) would allow checking whether $t \in
\mathit{ind}_{S_\N}(o)$ for any $t$ and $o$ very efficiently, by exchanging
$O(log K)$ messages where $K \simeq n.$

\subsection{Supporting tacit name-based articulations}
\label{sec:DHT_T}

In a complementary way to the network inversion discussed in the previous
subsection, suppose that each element of $T_{\N}$ has a unique global identity and
meaning, \ie~if the taxonomies of two peers $\ES_1$ and $\ES_2$ contain two terms
having the same name, say $\mathtt{train}_1$ and $\mathtt{train}_2$, then these
two correspond to the same ``concept'' $\mathtt{train}$.  Making the above
assumption means that $T_{\N}$ exists before the formation of the network and
that $T_{\N}$ comprises elements that have the same meaning for all sources that
will form the network\footnote{ In other words, it is assumed that there is
  already a set of agreements between all peers on a common vocabulary.  These
  agreements are not represented explicitly within the network (they are
  external).  }, \eg~ $T_{\N}$ could be the set of all Greek words, or all terms
of the CACM taxonomy.  Note that structured P2P systems (like Chord and CAN) are
based on this assumption (\ie~that there is a globally accepted set of keys).
In contrast, our model considers that if the same term (e.g. word) appears in
the taxonomies of two different peers, then these occurrences do not denote the
same concept; for example $\mathtt{train}_1$ could mean ``wagon train'', while
$\mathtt{train}_2$ could mean ``instruct''.  So in our model all agreements
should be represented explicitly in articulations.

However, we could extend our model so that to be able to also capture a
preexisting globally accepted terminology $T_{\N}$, as follows: If a term $t$
appears in two peers $\ES_1$ and $\ES_2$, then we could assume that $\ES_1$ has in its
articulation the relationship $t_2 \preceq t_1$, and that $\ES_2$ has in its
articulation the relationship $t_1 \preceq t_2$.  Note that this would result in
symmetric articulations, \ie~it is like assuming that we have one two-way
articulation $t_1 \sim t_2$ (that is known by both $\ES_1$ and $\ES_2$).  Although
we could capture in this way the existence of a globally accepted terminology
$T_{\N}$, in practice the definition of articulations would be problematic:
how could a peer discover that another peer uses the same term?

This problem could be solved by employing a DHT that stores the terms and the
addresses of the peers that use these terms.  Specifically, for each term $t$ in
$T_{\N}$ there will be one peer that stores the addresses of all peers that have
$t$ in their taxonomies.  It follows, that a peer can exploit the DHT in order
to get efficiently the implicit (term-to-term) articulations of its terms
(without having to discover by itself the online peers that happen to use terms
that it uses too).

Specifically, if $t$ is a term of a peer $P$ and $t$ is involved
in the query evaluation procedure (that takes place in $P$), then
$P$ should ask  the DHT
in order to get that addresses of the peers that also use $t$.
It follows that the
 calls to the DHT should be issued
in the context of the \ask~ procedure,
so as the resulting terms to be
taken into account as articulation hyperedges.
For example, if $\{1,3,5\}$ is the set of addresses returned by the DHT,
then
the peer behaves
as if  its articulation contained
the relationships
$t_1 \preceq t$,
$t_3 \preceq t$,
$t_5 \preceq t$.

Also note that a special prefix could
be used for
discriminating global terms
from non global terms, e.g.  $\mathtt{global:train}$.  This could be
extended to support several name spaces (e.g. $\mathtt{transportation:train}$,
$\mathtt{education:train}$).

Overall, we can exploit a DHT of this kind in order to support efficient query
evaluation in cases where both implicitly defined articulations (e.g.
name-based) and explicitly defined articulations (like those discussed in this
paper) are desired.

\section{Related work}
\label{sec:rw}

In this paper we studied the problem of evaluating content-based retrieval
queries in an entirely pure P2P architecture (without any form of structuring),
where each peer can have its own conceptual model expressed as a taxonomy.

To evaluate a query $q$ posed to a peer $\ES$, peer $\ES$ propagates the incoming
query (which is always expressed over its own taxonomy) only to those peers to
which $\ES$ has an articulation and who can contribute to the answer of the query
(the latter is determined by the taxonomy and the articulations of $\ES$).
Specifically, $\ES$ does not propagate the original query $q$, but a set of
queries each one expressed in the query language (here vocabulary) of the
recipient peer. Note that there is not any form of centralized index (like in
Napster \cite{Napster01}), nor any flooding of queries (like in Gnutella
\cite{Gnutella}), nor any form of partitioned global index (like in Chord
\cite{Stoica01} and CAN \cite{CAN01}). Instead we have a query propagation
mechanism that is query and articulation dependent (note that Semantic Overlay
Networks \cite{SONS2002} is a very simplistic approach to this). In case the
objects of the domain happen to have a unique global identity (like URI), then
automatic techniques can be applied for the construction of articulations (e.g.
see \cite{TzitzikasMeghiniCIA03}), and we can also obtain more rich object
descriptions by aggregating the descriptions that have been associated to each
object.

Moreover note that the peers of our model are quite autonomous in the sense that
they do not have to share or publish their stored objects, taxonomies or
mappings with the rest of the peers (neither to one central server, nor to the
on-line peers). To participate in the network, a peer just has to answer the
incoming queries by using its local base, and to propagate queries to those
peers that according to its ``knowledge" (i.e.  taxonomy + articulations) may
contribute to the evaluation of the query. However both of the above tasks are
optional and at the ``will" of the peer.

The literature about information integration distinguishes two main approaches:
the {\em local-as-view} (LAV) and the {\em global-as-view} (GAV) approach (see
\cite{SWWS-2001,LenzeriniPODS02} for a comparison).  In the LAV approach the
contents of the sources are defined as views over the mediator's schema, while
in the GAV approach the mediator's virtual contents are defined as views of the
contents of the sources.  The former approach offers flexibility in representing
the contents of the sources, but query answering is ``hard" because this
requires answering queries using views (\cite{Duschka97b,Levy01VLDB,Ullman97}).  
On the other hand, the GAV approach offers easy query
answering (expansion of queries until getting to source relations), but the
addition/deletion of a source implies updating the mediator view, i.e. the
definition of the mediator relations.  In our case, and if the articulations
contain relationships between single terms, then we have the benefits of both
GAV and LAV approaches, \ie~(a) the query processing simplicity of the GAV
approach, as query processing basically reduces to unfolding the query using the
definitions specified in the mapping, so as to translate the query in terms of
accesses (\ie~queries) to the sources, and (b) the modeling scalability of the
LAV approach, \ie~the addition of a new underlying source does not require
changing the previous mappings.  On the other hand, term-to-query articulations
resemble the GAV approach.  In a P2P setting, the cycles create more complex
emergent relationships.  For example suppose a peer $A$ having an articulation
$b_1 \wedge b_2 \leq a_1$ to a peer $B$ (this is a GAV definition for $a_1$ of
$A$) and a peer $B$ having an articulation $a_1 \wedge a_2 \leq b_3$ to the peer
$A$ (this is a GAV definition for $b_3$ of $B$).  However by taking into account
the entire network, we result in the ``mixed'' relationship $b_1 \wedge b_2
\wedge a_2 \leq b_3$.

Recently, there have been several works on P2P systems endowed with logic-based models of the peers' information bases
and of the mappings relating them (called {\em P2P mappings}). These works can be classified in 2 broad categories: (1) those
assuming propositional or Horn clauses as representation language or as a computational framework, and (2) those based
on more powerful formalisms.  With respect to the former category (\eg, see~\cite{rousset}), our work makes an important
contribution, by providing a much simpler algorithm for performing query answering than those based on resolution.
Indeed, we do rely on the theory of propositional Horn clauses, but only for proving the correctness of our algorithm.
For implementing query evaluation, we devise an algorithm that avoids the (unnecessary) algorithmic complications that
plague the methods based on resolution. As an example, after appropriate transformations our framework can be seen as a
special case of that in~\cite{rousset}. Then, query evaluation can be performed by first computing the prime implicates
of the negation of each term in the query, using the resolution-based algorithms presented in~\cite{rousset}. As the
complexity of this problem is exponential w.r.t the size of the taxonomy and polynomial w.r.t. the size of $\obj,$ there
is no computational gain in using this approach. Instead, there is an algorithmic loss, since the method is much more
complicated than ours.

As for the second category above, works in this area have focused on providing highly expressive knowledge
representation languages in order to capture  the widest range of applications. Notably, \cite{CalvanesePODS04}
proposes a model allowing, among other things, for existential quantification both in the bodies and in the heads of the
mapping rules.  Inevitably, such languages pose computational problems: deciding membership of a tuple in the answer of
a query is undecidable in the framework proposed by~\cite{CalvanesePODS04}, while disjunction in the rules' heads makes
the same problem coNP-hard already for datalog with unary predicate (\ie~terms), as we have proved in
Section~\ref{sec:q2qdl}.  These problems are circumvented in both approaches by changing the semantics of a P2P network,
in particular by adopting an epistemic reading of mappings. 

Below, we review in more detail
several works dealing with the problem of answering (union of) conjunctive
queries posed to a peer in logic-based P2P frameworks.

In \cite{CalvaneseDBISP2P-2003}, a query answering algorithm for
simple P2P systems is presented where each peer $\ES$ is associated
with a local database, an (exported) peer schema, and a set of local
mapping rules from the schema of the local database to the peer
schema. P2P mapping rules are of the form $cq_1 \leadsto cq_2$, where
$cq_1, cq_2$ are conjunctive queries of the same arity $n \geq 1$
(possibly involving existential variables), expressed over the union
of the schemas of the peers, and over the schema of a single peer,
respectively\footnote{Note that P2P mapping rules of this kind can
  accommodate both GAV and LAV-style mappings, and are referred in the
  literature as GLAV mappings.}.  Note that this representation
framework partially subsumes our network source framework, since in
our case $cq_1, cq_2$ are of arity 1, $cq_1$ is a conjunctive query of
the form $u_1(x) \wedge ... \wedge u_r(x)$ over the terminology of a
single peer\footnote{Recall that this restriction can be easily
  relaxed.}  and $q_2$ is a single atom query $t(x)$ over the
terminology of the peer that the mapping (articulation) belongs to.
However, simple P2P systems cannot express the local to a peer $\ES$
taxonomy $\preceq_\ES$ of our framework.  Query answering in simple
P2P systems according to the first-order logic (FOL) semantics is in
general undecidable. Therefore, the authors adopt a new semantics
based on epistemic logic in order to get decidability for query
answering.  Notably, the FOL semantics and epistemic logic semantics
for our framework coincide.  In particular, in
\cite{CalvaneseDBISP2P-2003}, a centralized bottom-up algorithm is
presented which essentially constructs a finite database $RDB$ which
constitutes a ``representative" of all the epistemic models of the P2P
system.  The answers to a conjunctive query $q$ are the answers of $q$
w.r.t. $RDB$.  However, though this algorithm has polynomial time
complexity, it is centralized and it suffers from the drawbacks of
bottom-up computation that does not take into account the structure of
the query.

The work in \cite{CalvaneseDBISP2P-2003} is extended in
\cite{CalvanesePODS04}, where a more general framework for P2P systems
is considered, which fully subsumes our framework and whose semantics
is based on epistemic logic.  In particular, in
\cite{CalvanesePODS04}, a peer is also associated with a set of
(function-free) FOL formulas over the schema of the peer. A top-down
distributed query answering algorithm is presented which is based on
synchronous messaging. Essentially, the algorithm returns to the peer
where the original query is posed, a datalog program by transferring
the full extensions of the relevant to the query, peer source
predicates along the paths of peers involved in query processing. The
returned datalog program is used for providing the answers to the
query. Obviously, our algorithm has computational advantages w.r.t.
the algorithm in \cite{CalvanesePODS04}, since during query evaluation
only the full or partial answer to a term (sub)query is transfered to
the peer that posed the (sub)query, and not the full extensions of all
terms involved in its evaluation.

The framework in \cite{SerafChid00}, extends our framework by
considering (i) $n$-ary (instead of unary) predicates (\ie~P2P
mappings are general datalog rules) and (ii) a set of domain relations
(also suggested in \cite{SeGiMyBe03}), mapping the objects of one peer
to the objects of another peer. A distributed query answering
algorithm is presented based on synchronous messaging. However, the
algorithm will perform poorly in our restricted framework\footnote{In
  our framework, domain relations correspond to the identity
  relation.}, since when a peer receives a (sub)query, it iterates
through the relevant P2P mappings and for each one of them, sends a
(sub)query to the appropriate peer (waiting for its answer), until
fixpoint is reached.  In our case, when a peer receives a (sub)query,
each relevant P2P mapping is considered just once and no iteration
until fixpoint is required.

A P2P framework similar to \cite{CalvaneseDBISP2P-2003} is presented
in \cite{Halevy03a}, where query answering according to FOL semantics
is investigated.  Since in general, query answering is undecidable,
the authors present a centralized algorithm (employed in the Piazza
system \cite{HalevyTKDE04}), which however is complete (the algorithm
is always sound), only for the case that polynomial time complexity in
query answering can be achieved. This includes the condition that
inclusion P2P mappings are acyclic.  However, such a condition
severely restricts the modularity of the system. Note that our
algorithm is sound and complete even in the case that there are cycles
in the term dependency path and it always terminates. Thus, our
framework allows placing articulations between peers without further
checks. This is quite important, because the actual interconnections
are not under the control of any actor in the system.

In \cite{FranconiKLZ04a,FranconiKLZ04b}, the authors consider a
framework where each peer is associated with a relational database,
and P2P mapping rules contain conjunctive queries in both the head and
the body of the rule (possibly with existential variables), each
expressed over the alphabet of a single peer. Again the semantics of
the system is defined based on epistemic logic \cite{FranconiKLS03c}.
In these papers, a peer database update algorithm is provided allowing
for subsequent peer queries to be answered locally without fetching
data from other nodes at query time. The algorithm (which is based on
asynchronous messaging) starts at the peer which sends queries to all
neighbour peers according to the involved mapping rules. When a peer
receives a query, the query is processed locally by the peer itself
using its own data. This first answer is immediately replied back to
the node which issued the query and sub-queries are propagated
similarly to all neighbour peers. When a peer receives an answer, (i)
it stores the answer locally, (ii) it materializes the view
represented in the head on the involved mapping rule, and (ii) it
propagates the result to the peer that issued the (sub)query.  Answer
propagation stops when no new answer tuples are coming to the peer
through any dependency path, that is until fixpoint is reached. In our
case, the database update problem for a peer $\ES$ amounts to invoking
$\ES':\query(q)$ for each articulation $q \preceq t$ from $\ES$ to
another peer $\ES'$ and storing the answer locally to $\ES$.  Note
that our query answering algorithm is also based on asynchronous
messaging.  However, since it considers a limited framework, it is
much simpler and no computation until fixpoint is required. In
particular, for each term (sub)query issued to a peer through $\ask$,
only one answer is returned through $\tell$.

\section{Conclusions}
\label{sec:conc}

This study presents a model of a P2P network consisting of sources based on
taxonomies. A taxonomy states subsumption relationships between negation-free
DNF formulas on terms and negation-free conjunctions of terms. The language for
querying such sources offers Boolean combinations of terms, in which negation
can be efficiently handled by adopting a closed-world reading of the
information. An efficient, hypergraph-based query evaluation method is presented
for such sources, resting on results coming from the theory of propositional
clauses. It is also shown that extending the expressive power of the taxonomy
language by adding negation or full disjunction, leads to the intractability of
the decision problem.

A model of a P2P network, having sources as nodes, is subsequently presented.
The essential feature of the model is the possibility of relating the assumed
disjoint peer terminologies by means of subsumption relationships of the same
type as those in the taxonomies of the sources. The resulting system subscribes
to the universally accepted notion of P2P information system, recently
postulated also in the context of the so-called emergent
semantics~\cite{AbererEmergentSemantics04}. It is also shown that the results presented
in the paper do apply also if the subsumption relationships are formed by
arbitrarily mixing terms from different terminologies.

An efficient query evaluation procedure for queries stated against such a
network is presented, and proved correct. The procedure is a distributed version
of the centralized procedure, based on an asynchronous, message-based
interaction amongst the peers aimed at favoring scalability. Some optimization
techniques are also discussed, namely one based on caching, for which the
algorithms for message processing are given.

Finally, the work is related to the most relevant papers in the area of P2P
systems. It remains to be seen, whether the same efficiency can be obtained by allowing
full datalog as a representation language for information sources and for
articulations. Yet, it is evident that the
B-graph based algorithm presented in this paper does not extend immediately to
the general datalog case, due to the presence of multiple variables in the rules
and unification.
\small

\subsection*{Acknowledgments}
\label{ack}

We thank Nicolas Spyratos for inspiring this work. We also thank ERCIM for
offering Yannis Tzitzikas a fellowship in the course of which the work was
started, and the DELOS Network of Excellence Exchange Programme, for supporting
Carlo Meghini's visit to the Institute of Computer Science of FORTH, in the
course of which the work was completed.


\begin{thebibliography}{10}

\bibitem{Gnutella}
Gnutella(http://gnutella.wego.com).

\bibitem{Kazaa}
Kazaa (http://www.kazaa.com/).

\bibitem{Napster01}
Napster (www.naptster.com), 2001.

\bibitem{AbererEmergentSemantics04}
K.~Aberer, T.~Catarci, P.~Cudr{\'e}-Mauroux, T.~S. Dillon, S.~Grimm, M.~Hacid,
  A.~Illarramendi, M.~Jarrar, V.~Kashyap, M.~Mecella, E.~Mena, E.~J. Neuhold,
  A.~M. Ouksel, T.~Risse, M.~Scannapieco, F.~Saltor, L.~De Santis,
  S.~Spaccapietra, S.~Staab, R.~Studer, and O.~De Troyer.
\newblock {``Emergent Semantics Systems"}.
\newblock In {\em Procs. of the 1st Intern. IFIP Conference on Semantics of a
  Networked World (ICSNW 2004)}, pages 14--43, 2004.

\bibitem{AbHV95}
S.~Abitebul, R.~Hull, and V.~Vianu.
\newblock {\em Foundations of Databases}.
\newblock Addison-Wesley, 1995.
\newblock ISBN: 0-201-53771-0.

\bibitem{rousset}
Philippe Adjiman, Philippe Chatalic, Francois Goasdou{\'e}, Marie-Christine
  Rousset, and Laurent Simon.
\newblock Distributed reasoning in a peer-to-peer setting: Application to the
  semantic web.
\newblock {\em Journal of Artificial Intelligence Research}, 25:269--314, 2006.

\bibitem{Bernstein02}
Philip~A. Bernstein, F.~Giunchiglia, A.~Kementsietsidis, J.~Mylopoulos,
  L.~Serafini, and I.~Zaihrayeu.
\newblock {``Data Management for Peer-to-Peer Computing: A Vision"}.
\newblock In {\em Proceedings of WebDB02}, Madison, Wisconsin, June 2002.

\bibitem{Bolosky00}
W.~J. Bolosky, J.~R. Douceur, D.~Ely, and M.~Theimer.
\newblock {``Feasibility of a Serveless Distributed File System Deployed on an
  Existing Set of Desktop PCs"}.
\newblock In {\em Proceedings of Measurement and Modeling of Computer Systems},
  June 2000.

\bibitem{CalvaneseDBISP2P-2003}
Diego Calvanese, Elio Damaggio, Giuseppe De~Giacomo, Maurizio Lenzerini, and
  Riccardo Rosati.
\newblock {``Semantic Data Integration in P2P Systems"}.
\newblock In {\em Procs. of the First International Workshop on Databases,
  Information Systems and Peer-to-Peer Computing (DBISP2P 2003)}, pages 79--90,
  2003.

\bibitem{SWWS-2001}
Diego Calvanese, Giuseppe De~Giacomo, and Maurizio Lenzerini.
\newblock {``A Framework for Ontology Integration"}.
\newblock In {\em Proc. of the 2001 Int. Semantic Web Working Symposium (SWWS
  2001)}, pages 303--316, Stanford University, California, USA, July 30 -
  August 1 2001.

\bibitem{CalvanesePODS04}
Diego Calvanese, Giuseppe~De Giacomo, Maurizio Lenzerini, and Riccardo Rosati.
\newblock {``Logical foundations of peer-to-peer data integration"}.
\newblock In {\em Procs. of the 23rd ACM symposium on Principles of database
  systems, PODS'2004}, pages 241--251, New York, NY, USA, 2004. ACM Press.

\bibitem{gct1990}
S.~Ceri, G.~Gottlob, and L.~Tanca.
\newblock {\em Logic Programming and Databases}.
\newblock Springer Verlag, 1990.

\bibitem{SONS2002}
Arturo Crespo and Hector Garcia-Molina.
\newblock {``Semantic Overlay Networks for P2P Systems"}.
\newblock Technical report, Computer Science Department, Stanford University,
  October 2002.

\bibitem{planetP2003}
Francisco~Matias Cuenca-Acuna, Christopher Peery, Richard~P. Martin, and Thu~D.
  Nguyen.
\newblock {``PlanetP: Using Gossiping to Build Content Addressable Peer-to-Peer
  Information Sharing Communities"}.
\newblock In {\em Procs. of 12th IEEE International Symposium on High
  Performance Distributed Computing (HPDC-12)}, pages 236--249. IEEE Press,
  June 2003.

\bibitem{degv01}
E.~Dantsin, T.~H. Eiter, G.~Gottlob, and A.~Voronkov.
\newblock {``Complexity and Expressive Power of Logic Programming"}.
\newblock {\em ACM Computing Survey}, ACM Computing Survey 33(3):374--425,
  September 2001.

\bibitem{Duschka97b}
Oliver~M. Duschka and Michael~R. Genesereth.
\newblock Answering recursive queries using views.
\newblock In {\em Procs. of the 16th ACM SIGACT-SIGMOD-SIGART Symposium on
  Principles of Database Systems (PODS'97)}, pages 109--116, Tucson, Arizona,
  12-14 May 1997.

\bibitem{fejer}
P.A. Fejer and D.A. Simovici.
\newblock {\em Mathematical Foundations of Computer Science. Volume 1: Sets,
  Relations, and Induction}.
\newblock Springer-Verlag, 1991.

\bibitem{FranconiKLS03c}
E.~Franconi, G.~M. Kuper, A.~Lopatenko, and L.~Serafini.
\newblock {``A Robust Logical and Computational Characterisation of
  Peer-to-Peer Database Systems"}.
\newblock In {\em Procs. of the First International Workshop on Databases,
  Information Systems, and Peer-to-Peer Computing (DBISP2P 2003)}, pages
  64--76, 2003.

\bibitem{FranconiKLZ04b}
E.~Franconi, G.~M. Kuper, A.~Lopatenko, and I.~Zaihrayeu.
\newblock {``A Distributed Algorithm for Robust Data Sharing and Updates in P2P
  Database Networks"}.
\newblock In {\em Procs. of the EDBT'04 Intern. Workshop on Peer-to-Peer
  Computing and Databases (P2P\&DB 2004)}, pages 446--455, 2004.

\bibitem{FranconiKLZ04a}
E.~Franconi, G.~M. Kuper, A.~Lopatenko, and I.~Zaihrayeu.
\newblock {``Queries and Updates in the coDB Peer to Peer Database System"}.
\newblock In {\em Procs. of the 30th International Conference on Very Large
  Data Bases (VLDB 2004)}, pages 1277--1280, 2004.

\bibitem{gallo93}
Giorgio Gallo, Giustino Longo, and Stefano Pallottino.
\newblock {``Directed Hypergraphs and Applications"}.
\newblock {\em Discrete Applied Mathematics}, 42(2):177--201, 1993.

\bibitem{Ganter99}
Bernhard Ganter and Rudolf Wille.
\newblock {\em {``Formal Concept Analysis: Mathematical Foundations"}}.
\newblock Springer-Verlag, Heidelberg, 1999.

\bibitem{HalevyTKDE04}
A.~Y. Halevy, Z.~G. Ives, J.~Madhavan, P.~Mork, D.~Suciu, and I.~Tatarinov.
\newblock {``The Piazza Peer Data Management System"}.
\newblock {\em IEEE Transactions on Knowledge and Data Engineering},
  16(7):787--798, 2004.

\bibitem{Halevy03b}
Alon Halevy, Zachary Ives, Peter Mork, and Igor Tatarinov.
\newblock {``Piazza: Data Management Infrastructure for Semantic Web
  Applications"}.
\newblock In {\em Procs. of the 12th International Conference on World Wide Web
  (WWW 2003)}, pages 556 -- 567, May 2003.

\bibitem{Halevy03a}
Alon Halevy, Zachary Ives, Dan Suciu, and Igor Tatarinov.
\newblock {``Schema Mediation in Peer Data Management Systems"}.
\newblock In {\em Procs. of the 19th International Conference on Data
  Engineering (ICDE'03)}, pages 505--518, March 2003.

\bibitem{Levy01VLDB}
Alon~Y. Halevy.
\newblock {``Answering Queries Using Views: A Survey"}.
\newblock {\em VLDB Journal}, 10(4):270--294, 2001.

\bibitem{Koubarakis02}
M.~Koubarakis and C.~Tryfonopoulos.
\newblock {``Peer-to-Peer Agent Systems for Textual Information Dissemination:
  Algorithms and Complexity"}.
\newblock In {\em Proceedings of the UK Workshop on Multiagent
  Systems,UKMAS'02}, Liverpool, UK, 2002.

\bibitem{LenzeriniPODS02}
Maurizio Lenzerini.
\newblock {``Data Integration: A Theoretical Perspective"}.
\newblock In {\em Proc. of the 21st ACM SIGMOD-SIGACT-SIGART Symposium on
  Principles of Database Systems (PODS 2002)}, pages 233--246, Madison,
  Wisconsin, USA, June 2002.

\bibitem{LingWISE02}
Bo~Ling, Zhiguo Lu, Wee~Siong Ng, BengChin Ooi, Kian-Lee Tan, and Aoying Zhou.
\newblock {``A Content-Based Resource Location Mechanism in PeerIS"}.
\newblock In {\em Proceedings of the 3rd International Conference on Web
  Information Systems Engineering, WISE 2002}, Singapore, December 2002.

\bibitem{ArticulatedODBASE04}
Carlo Meghini and Yannis Tzitzikas.
\newblock {``Querying Articulated Sources"}.
\newblock In {\em Procs. of the 3rd Intern. Conference on Ontologies, Databases
  and Applications of Semantics for Large Scale Information Systems,
  ODBASE'2004}, pages 945--962, Larnaca, Cyprus, October 2004.

\bibitem{Edutella02b}
W.~Nejdl, B.~Wolf, C.~Qu, S.~Decker, M.~Sintek, A.~Naeve, M.~Nilsson,
  M.~Palmer, and T.~Risch.
\newblock {''EDUTELLA: A P2P networking infrastructure based on RDF"}.
\newblock In {\em Procs. of the 11th International Conference on World Wide Web
  (WWW'02)}, pages 604 -- 615, 2002.

\bibitem{Edutella02}
W.~Nejdl, B.~Wolf, S.~Staab, and J.~Tane.
\newblock {``EDUTELLA: Searching and Annotating Resources within an RDF-based
  P2P Network"}.
\newblock In {\em Procs of the WWW2002 International Workshop on the Semantic
  Web}, Honolulu, Havaii, May 2002.

\bibitem{CAN01}
Sylvia Ratnasamy, Paul Francis, Mark Handley, Richard Karp, and Scott Shenker.
\newblock {``A Scalable Content Addressable Network"}.
\newblock In {\em Procs. of 2001 Conference on Applications, Technologies,
  Architectures, and Protocols for Computer Communications (SIGCOMM'2001)},
  pages 161 -- 172, 2001.

\bibitem{SerafChid00}
L.~Serafini and C.~Ghidini.
\newblock {``Using Wrapper Agents to Answer Queries in Distributed Information
  Systems"}.
\newblock In {\em Procs. of the First International Conference on Advances in
  Information Systems (ADVIS '00)}, pages 331--340. Springer-Verlag, 2000.

\bibitem{SeGiMyBe03}
L.~Serafini, F.~Giunchiglia, J.~Mylopoulos, and P.~A. Bernstein.
\newblock {``Local Relational Model: A Logical Formalization of Database
  Coordination"}.
\newblock In {\em Procs. of the 4th International and Interdisciplinary
  Conference on Modeling and Using Context(CONTEXT 2003)}, pages 286--299,
  2003.

\bibitem{Stoica01}
Ion Stoica, Robert Morris, David Karger, M.~Frans Kaashoek, and Hari
  Balakrishnan.
\newblock {``Chord: A Scalable Peer-to-peer Lookup Service for Internet
  Applications"}.
\newblock In {\em Proceedings of the 2001 ACM SIGCOMM Conference}, 2001.

\bibitem{pSearch2002}
Chunqiang Tang, Zhichen Xu, and Mallik Mahalingam.
\newblock {``pSearch: Information Retrieval in Structured Overlays"}.
\newblock In {\em Procs. of ACM HotNets-I}, October 2002.

\bibitem{TzitzikasMeghiniCIA03}
Yannis Tzitzikas and Carlo Meghini.
\newblock {``Ostensive Automatic Schema Mapping for Taxonomy-based Peer-to-Peer
  Systems"}.
\newblock In {\em Seventh International Workshop on Cooperative Information
  Agents, CIA-2003}, pages 78--92, Helsinki, Finland, August 2003.
\newblock (Best Paper Award).

\bibitem{TzitzikasMeghiniCoopIS03}
Yannis Tzitzikas and Carlo Meghini.
\newblock {``Query Evaluation in Peer-to-Peer Networks of Taxonomy-based
  Sources"}.
\newblock In {\em Procs. of 19th Int. Conf. on Cooperative Information Systems,
  CoopIS'2003}, pages 263--281, Catania, Sicily, Italy, November 2003.

\bibitem{TzitzikasMeghiniER03}
Yannis Tzitzikas, Carlo Meghini, and Nicolas Spyratos.
\newblock {``Taxonomy-based Conceptual Modeling for Peer-to-Peer Networks"}.
\newblock In {\em Procs. of 22th Int. Conf. on Conceptual Modeling, ER'2003},
  pages 446--460, Chicago, Illinois, October 2003.

\bibitem{Ullman88I}
Jeffrey~D. Ullman.
\newblock {\em {``Principles of Database and Knowledge-Base Systems, Vol. I"}}.
\newblock Computer Science Press, 1988.

\bibitem{Ullman97}
Jeffrey~D. Ullman.
\newblock {``Information integration using logical views"}.
\newblock In {\em Procs. of the 6th Int. Conf. on Database Theory (ICDT-97)},
  pages 19--40, Delphi, Greece, 8-10 January 1997.

\end{thebibliography}

\appendix

\section{Completion of the example}
\label{sec:exa}

We resume the example from the processing of the message
$P_a$:\tell($q3,I(a3)$).
\begin{itemize}
\item $P_a$:\tell($q3,I(a3)$)

  \tell~ finds the object in the log and updates it. The old log on $P_a$ was:

  \begin{tabular}{l}
    $P_a$ log \\ \hline
    ($P_a,q1,t,2,\{\{q2,q3\}\}$)
  \end{tabular}

  The new log is:

  \begin{tabular}{l}
    $P_a$ log \\ \hline
    ($P_a,q1,t,1,\{\{q2,I(a3)\}\}$)
  \end{tabular}

\item $P_a$:\ask($P_a,q2,a2,\{t,a2\}$)

    Since there are two incoming hyperedges in $a2,$ both in $P_b,$ \ask~enqueues 3 \ask~messages
    to $P_b,$ one for each involved term:
    \begin{itemize}
    \item $P_b$:\ask($P_a,q4,b3,\{t,a2,b3\}$)
    \item $P_b$:\ask($P_a,q5,b1,\{t,a2,b1\}$)
    \item $P_b$:\ask($P_a,q6,b2,\{t,a2,b2\}$)
    \end{itemize}
    It then persists the corresponding log object. The new log is:

  \begin{tabular}{l}
    $P_a$ log \\ \hline
    ($P_a,q1,t,1,\{\{q2,I(a3)\}\}$) \\
    ($P_a,q2,a2,3,\{\{q4\},\{q5,q6\}\}$)
  \end{tabular}

  and issues the 3 enqueued messages.

  \item $P_b$:\ask($P_a,q4,b3,\{t,a2,b3\}$)

  Since there are no incoming hyperedges in $b3,$ the message
  $P_a$:\tell($q4,I(b3)$) is produced.

\item $P_a$:\tell($q4,I(b3)$)

  \tell~ finds the object in the log and updates it. The updated log is:

  \begin{tabular}{l}
    $P_a$ \\ \hline
    ($P_a,q1,t,1,\{\{q2,I(a3)\}\}$) \\
    ($P_a,q2,a2,2,\{\{I(b3)\},\{q5,q6\}\}$)
  \end{tabular}

\item  $P_b$:\ask($P_a,q5,b1,\{t,a2,b1\}$)

  Since there are two incoming hyperedges in $b1,$ \ask~enqueues 2 \ask~messages to $P_c,$ one for each involved term:
    \begin{itemize}
    \item $P_c$:\ask($P_b,q7,c1,\{t,a2,b1,c1\}$)
    \item $P_c$:\ask($P_b,q8,c2,\{t,a2,b1,c2\}$)
    \end{itemize}
  It then persists the corresponding log object. The log is now:

  \begin{tabular}{l}
    $P_b$ log \\ \hline
    ($P_a,q5,b1,2,\{\{q7\},\{q8\}\}$)
  \end{tabular}

\item $P_b$:\ask($P_a,q6,b2,\{t,a2,b2\}$)

  Since there is one incoming hyperedge in $b2,$ \ask~enqueues 2 \ask~messages
  to $P_c,$ one for each involved term:
    \begin{itemize}
    \item $P_c$:\ask($P_b,q9,c2,\{t,a2,b2,c2\}$)
    \item $P_c$:\ask($P_b,q10,c3,\{t,a2,b2,c3\}$)
    \end{itemize}
  It then persists the corresponding log object. The log is now:

  \begin{tabular}{l}
    $P_b$ log \\ \hline
    ($P_a,q5,b1,2,\{\{q7\},\{q8\}\}$) \\
    ($P_a,q6,b2,2,\{\{q9,q10\}\}$)
  \end{tabular}

\item  $P_c$:\ask($P_b,q7,c1,\{t,a2,b1,c1\}$)

  Since there are no incoming hyperedges in $c1,$ \ask~generates
  $P_b$:\tell($q7,I(c1)$).

\item $P_b$:\tell($q7,I(c1)$)

  \tell~ finds the object in the log and updates it. The new log is:

  \begin{tabular}{l}
    $P_b$ log \\ \hline
    ($P_a,q5,b1,1,\{\{I(c1)\},\{q8\}\}$) \\
    ($P_a,q6,b2,2,\{\{q9,q10\}\}$)
  \end{tabular}

\item  $P_c$:\ask($P_b,q8,c2,\{t,a2,b1,c2\}$)

  Since there is one incoming hyperedge in $c2$ but its tail has a non-empty
  intersection with the set of visited terms, just a \tell~message
  results: $P_b$:\tell($q8,I(c2)$).

\item $P_b$:\tell($q8,I(c2)$)

  \tell~ finds the object in the log and updates it. The new log is:

  \begin{tabular}{l}
    $P_b$ log \\ \hline
    ($P_a,q5,b1,0,\{\{I(c1)\},\{I(c2)\}\}$) \\
    ($P_a,q6,b2,2,\{\{q9,q10\}\}$)
  \end{tabular}

  There are no more open calls in the updated log object, therefore the answer
  to the query $q5$ can be computed as $I(c1)\cup I(c2).$ Then the object is
  deleted permanently from the log and the message $P_a$:\tell($q5,I(b1)\cup
  I(c1)\cup I(c2)$) is issued.

\item  $P_a$:\tell($q5,I(b1)\cup I(c1)\cup I(c2)$)

  \tell~ finds the object in the log and updates it. The new log is:

  \begin{tabular}{l}
    $P_a$ log \\ \hline
    ($P_a,q1,t,1,\{\{q2,I(a3)\}\}$) \\
    ($P_a,q2,a2,1,\{\{I(b3)\},\{I(b1)\cup I(c1)\cup I(c2),q6\}\}$)
  \end{tabular}

\item  $P_c$:\ask($P_b,q9,c2,\{t,a2,b2,c2\}$)

  Since there is one incoming hyperedge in $c2,$ \ask~enqueues 2 \ask~messages
  to $P_b,$ one for each involved term:
    \begin{itemize}
    \item $P_b$:\ask($P_c,q11,b1,\{t,a2,b2,c2,b1\}$)
    \item $P_b$:\ask($P_c,q12,b3,\{t,a2,b2,c2,b3\}$)
    \end{itemize}
  It then persists the corresponding log object. The updated log is:

  \begin{tabular}{l}
    $P_c$ log \\ \hline
    ($P_b,q9,c2,2,\{\{q11,q12\}\}$)
  \end{tabular}

\item $P_c$:\ask($P_b,q10,c3,\{t,a2,b2,c3\}$)

  Since there are no incoming hyperedges in $c3$ a \tell~message results:
  $P_b$:\tell($q10,I(c3)$).

\item $P_b$:\tell($q10,I(c3)$)

  \tell~ finds the object in the log and updates it. The updated log is:

  \begin{tabular}{l}
    $P_b$ log \\ \hline
    ($P_a,q6,b2,1,\{\{q9,I(c3)\}\}$)
  \end{tabular}

\item $P_b$:\ask($P_c,q11,b1,\{t,a2,b2,c2,b1\}$)

  There are two incoming hyperedges in $b1,$ but the one having $c2$ in the tail
  generates no \ask~messages. The only \ask~enqueued is therefore:
    \begin{itemize}
    \item $P_c$:\ask($P_b,q13,c1,\{t,a2,b2,c2,b1,c1\}$)
    \end{itemize}
  It then persists the corresponding log object. The updated log is:

  \begin{tabular}{l}
    $P_b$ log \\ \hline
    ($P_a,q6,b2,1,\{\{q9,I(c3)\}\}$) \\
    ($P_c,q11,b1,1,\{\{q13\}\}$)
  \end{tabular}

\item $P_b$:\ask($P_c,q12,b3,\{t,a2,b2,c2,b3\}$)

  Since there are no incoming hyperedges in $b3,$ it results:
  $P_c$:\tell($q12,I(b3)$).

\item $P_c$:\tell($q12,I(b3)$)

  \tell~ finds the object in the log and updates it. The new log is:

  \begin{tabular}{l}
    $P_c$ log \\ \hline
    ($P_b,q9,c2,1,\{\{q11,I(b3)\}\}$)
  \end{tabular}

\item $P_c$:\ask($P_b,q13,c1,\{t,a2,b2,c2,b1,c1\}$)

  Since there are no incoming hyperedges in $c1,$ \ask~issues
  $P_b$:\tell($q13,I(c1)$).

\item $P_b$:\tell($q13,I(c1)$)

  \tell~ finds the object in the log and updates it. The new log is:

  \begin{tabular}{l}
    $P_b$ log \\ \hline
    ($P_a,q6,b2,1,\{\{q9,I(c3)\}\}$) \\
    ($P_c,q11,b1,0,\{\{I(c1)\}\}$)
  \end{tabular}

  There are no more open calls in the updated log object, therefore the answer
  to the query $q11$ can be computed as $I(c1).$ Then the object is permanently
  deleted from the log and the message $P_c$:\tell($q11,I(b1)\cup I(c1)$) is
  issued.

\item $P_c$:\tell($q11,I(b1)\cup I(c1)$)

  \tell~ finds the object in the log and updates it. The updated log is:

  \begin{tabular}{l}
    $P_c$ log \\ \hline
    ($P_b,q9,c2,0,\{\{I(b1)\cup I(c1),I(b3)\}\}$) \\
  \end{tabular}

  There are no more open calls in the updated log object, therefore the answer
  to the query $q9$ can be computed. Then the object is permanently deleted from
  the log and the message $P_b$:\tell($q9,[(I(b1)\cup I(c1))\cap I(b3)]\cup I(c2)$)
  is issued.

\item $P_b$:\tell($q9,[(I(b1)\cup I(c1))\cap I(b3)]\cup I(c2)$)

  \tell~ finds the object in the log and updates it. The updated log is:

  \begin{tabular}{l}
    $P_b$ log \\ \hline
    ($P_a,q6,b2,0,\{\{[(I(b1)\cup I(c1))\cap I(b3)]\cup I(c2),I(c3)\}\}$)
  \end{tabular}

  There are no more open calls in the updated log object, therefore the answer
  to the query $q6$ can be computed. Then the object is permanently deleted from
  the log and the message $P_a$:\tell($q6,[X\cap I(c3)]\cup I(b2)$) is issued,
  where
  \[
  X=[(I(b1)\cup I(c1))\cap I(b3)]\cup I(c2)
  \]

\item  $P_a$:\tell($q6,[X\cap I(c3)]\cup I(b2)$)

  \tell~ finds the object in the log and updates it. The updated log is:

  \begin{tabular}{l}
    $P_a$ log \\ \hline
    ($P_a,q1,t,1,\{\{q2,I(a3)\}\}$) \\
    ($P_a,q2,a2,0,\{\{I(b3)\},\{I(b1)\cup I(c1)\cup I(c2),[X\cap I(c3)]\cup I(b2)\}\}$)
  \end{tabular}

  There are no more open calls in the updated log object, therefore the answer
  to the query $q2$ can be computed. Then the object is permanently deleted from
  the log and the message $P_a$:\tell($q2,I(a2)\cup I(b3)\cup (Y\cap Z)$ is issued,
  where
  \begin{eqnarray*}
  Y & = & I(b1)\cup I(c1)\cup I(c2) \\
  Z & = & [X\cap I(c3)]\cup I(b2)
  \end{eqnarray*}

\item  $P_a$:\tell($q2,I(a2)\cup I(b3)\cup (Y\cap Z))$

  \tell~ finds the object in the log and updates it. The new log is:

  \begin{tabular}{l}
    $P_a$ log \\ \hline
    ($P_a,q1,t,0,\{\{I(a3)\cap (I(a2)\cup I(b3)\cup (Y\cap Z))\}\}$)
  \end{tabular}

  There are no more open calls in the updated object and $q1 \not \in T_{P_a}$.
  Therefore, $q1$ must be a user (external) query.

  The \query~procedure will realize that $q1$ is complete, and
  return the answer to the user, thus concluding query evaluation.

\end{itemize}

\end{document}